\documentclass[journal,twocolumn]{IEEEtran}
\usepackage[T1]{fontenc}
\usepackage{cite}
\usepackage{amsmath,amssymb}
\usepackage{bm}
\usepackage{graphicx}
\usepackage{subfigure}
\usepackage{textcomp}
\usepackage{xcolor}
\usepackage{threeparttable}
\usepackage{multirow}
\usepackage{verbatim}
\usepackage{url}
\usepackage{hyperref}
\usepackage{comment}
\usepackage{color}
\usepackage{booktabs}
\usepackage{amsthm}
\usepackage{makecell}
\usepackage{cuted}

\usepackage[linesnumbered, ruled]{algorithm2e}
\SetKwRepeat{Do}{do}{while}%

\def\BibTeX{{\rm B\kern-.05em{\sc i\kern-.025em b}\kern-.08em
    T\kern-.1667em\lower.7ex\hbox{E}\kern-.125emX}}
\begin{document}

\title{Deep Learning-based Low-Overhead Beam Alignment for mmWave Massive MIMO Systems}

\author{{Weijie Jin, \IEEEmembership{Student Member, IEEE}, Jing Zhang, \IEEEmembership{Member, IEEE}, Hengtao He, \IEEEmembership{Member, IEEE}, \\Chao-Kai Wen, \IEEEmembership{Fellow, IEEE}, Xiao Li, \IEEEmembership{Member, IEEE}, Shi Jin, \IEEEmembership{Fellow, IEEE}}

\thanks{W. Jin, J. Zhang, H. He, X. Li, and S. Jin are with the School of Information Science and Engineering, Southeast University, Nanjing, China (email: \{jinweijie, jingzhang, hehengtao, li\_xiao, jinshi\}@seu.edu.cn).

C.-K. Wen is with the Institute of Communications Engineering, National Sun Yat-sen University, Kaohsiung 80424, Taiwan (email: chaokai.wen@mail.nsysu.edu.tw).
}
\vspace{-20pt}
}
\maketitle

\begin{abstract}
Millimeter-wave massive multiple-input multiple-output systems employ highly directional beamforming to overcome severe path loss, and their performance critically depends on accurate beam alignment. Conventional codebook-based methods offer low training overhead but suffer from limited angular resolution and sensitivity to hardware impairments. To address these challenges, we propose a deep learning-enhanced super-resolution beam alignment framework with three key components. First, we design the Quaternary Search-based Super-Resolution (QSSR) algorithm, which leverages the monotonic power ratio property between two discrete Fourier transform (DFT) codebook beams to achieve super-resolution angle estimation without increasing measurement complexity relative to binary search. Second, we develop QSSR-Net, a gated recurrent unit-based neural network that exploits sequential multi-layer beam measurements to capture angular dependencies, thereby improving estimation accuracy, robustness to noise, and generalization across diverse propagation environments. Third, to mitigate the adverse effects of hardware impairments such as antenna position and phase errors, we propose a parametric self-calibration method that requires no additional hardware overhead and adapts compensation parameters in real time. Simulation results show that the proposed framework consistently outperforms binary search and even exhaustive search at high signal-to-noise ratios, achieving substantial performance gains while maintaining low overhead.
\end{abstract}


\section{Introduction}

\IEEEPARstart{M}{illimeter}-wave (mmWave) massive multiple-input multiple-output (MIMO) systems are pivotal for next-generation wireless communications, offering high data rates and enhanced spectral efficiency. However, the high carrier frequency leads to severe signal attenuation and distortion, necessitating highly directional beamforming for reliable transmission. Existing beamforming strategies are broadly categorized by their reliance on channel state information (CSI). CSI-based methods, while effective for optimizing metrics such as maximizing weighted sum rate \cite{zhaoRethinkingWMMSECan2023, jinModelDrivenDeepLearning2023,jinLowComplexityJointBeamforming2024} or minimizing transmit power \cite{liangDataModelDrivenDeep2024,xiaDeepLearningFramework2020}, suffer from scalability issues due to high channel estimation and feedback overhead. Furthermore, their underlying non-convex optimization problems often incur significant computational complexity. Given the prohibitive cost and power consumption of fully digital beamforming architectures at mmWave frequencies, analog-only or hybrid architectures are commonly employed \cite{alkhateebChannelEstimationHybrid2014, sohrabiHybridDigitalAnalog2016,xiaoHierarchicalCodebookDesign2016}. In such configurations, codebook-based methods are proposed as an effective solution \cite{ heSuboptimalBeamSearch2015, qiHierarchicalCodebookBasedMultiuser2020}. These methods avoid explicit channel estimation by performing beam alignment through beam training, wherein transmit and receive beams are sequentially selected from a predefined codebook, and the corresponding received signal strengths are measured to identify the optimal beam pairs for subsequent data transmission. These approaches offer lower complexity and competitive performance, making them attractive for practical massive MIMO deployments. 

{\bf Related Works}.  
As the number of antennas increases, the size of the codebook grows accordingly, making beam alignment increasingly challenging. To mitigate the prohibitive overhead of exhaustive search, the authors of \cite{heSuboptimalBeamSearch2015} proposed a hierarchical codebook structure, where wide beams are formed by selectively deactivating antenna elements and leveraging discrete Fourier transform (DFT) codebooks. Beam measurements at each layer guide the direction of subsequent narrower beams, effectively reducing the beam training overhead from linear to logarithmic with respect to the number of antennas. A similar concept was explored in \cite{xiaoHierarchicalCodebookDesign2016}, which divides the antenna array into subarrays and steers them toward spatially separated directions to generate broader beams. While these approaches significantly reduce beam training overhead, their beam alignment accuracy remains limited due to the underutilization of environmental statistical information.

Considering the advantages of deep learning (DL) in channel estimation \cite{yuAdaptiveRobustDeep2023} and signal detection \cite{heModeldrivenDeepLearning2020}, it has also emerged as a powerful tool for beam management \cite{linBsNetDeepLearningBased2019,fanSuperResolutionBasedBeam2022,zhangReinforcementLearningBeam2022,maDeepLearningAssisted2021,maMachineLearningBeam2020,wuFastMmwaveBeam2019}, aiming to enhance beam alignment accuracy and reduce the number of required measurements. In \cite{linBsNetDeepLearningBased2019}, beam alignment was first formulated as an image reconstruction problem, where the full set of narrow-beam measurements is recovered from a limited number of observations, and environmental statistical information is implicitly learned by the neural network. Building on this, \cite{fanSuperResolutionBasedBeam2022} proposed inferring the narrow-beam response from wide-beam measurements, demonstrating superior performance over image reconstruction methods. In \cite{zhangReinforcementLearningBeam2022}, a deep reinforcement learning framework was introduced to adaptively optimize beam patterns through interaction with the environment, using only received signal power as feedback. However, these approaches largely overlook the structure of codebooks and treat the beam measurement as an image, thereby neglecting intrinsic correlations among beams. This reduces model robustness and leads to performance degradation under environmental changes.

Researchers have also exploited multimodal information to improve beam alignment performance, often leveraging DL to map inputs such as user location, camera imagery, and LiDAR point clouds to optimal beam indices \cite{hengMachineLearningAssistedBeam2021, salehiDeepLearningMultimodal2022, salehiTextOmniCNNModalityAgnosticNeural2024}. However, they typically require additional hardware and feedback mechanisms, thus increasing system complexity and deployment costs. In contrast, another line of research focuses on using DL to learn site-specific probing beams that extract maximal channel information from fewer measurements \cite{yangHierarchicalBeamAlignment2024, myersDeepLearningBasedBeam2020, hengLearningSiteSpecificProbing2022, hengGridFreeMIMOBeam2024, hengSiteSpecificBeamAlignment2024}. These approaches include learning probing beams \cite{hengLearningSiteSpecificProbing2022, hengGridFreeMIMOBeam2024}, developing hierarchical frameworks for optimal beam estimation \cite{yangHierarchicalBeamAlignment2024}, or designing customized compressed sensing matrices \cite{myersDeepLearningBasedBeam2020}. Despite achieving significant performance gains, these approaches heavily rely on site-specific training data, which limits their generalizability across diverse environments.

Among DL-based methods, the super-resolution approach proposed in \cite{hengGridFreeMIMOBeam2024} represents a notable departure from conventional codebook-based beam alignment. Instead of selecting beams from a quantized codebook, it directly synthesizes beams in the continuous angular domain using measurements from learned site-specific probing beams. Despite its promising performance, the resulting model is highly environment dependent, requiring retraining when the environments change, which incurs substantial overhead and limits its practicality in dynamic scenarios. Super-resolution beam alignment has also been investigated in \cite{wangSuperResolutionWideBeamTraining2024}, which further demonstrates the potential performance gains achievable with super-resolution techniques. However, although this approach avoids learning-based training, it suffers from prohibitively high beam measurement overhead as the antenna array size increases. Moreover, it lacks a systematic analysis of the relationship among the received powers of intra-layer beams, thereby constraining its scalability and limiting deeper theoretical insight.

Prior studies often neglect hardware impairments in real-world antenna arrays, such as manufacturing imperfections and mutual coupling, which distort beam patterns and degrade performance. Antenna calibration is crucial to mitigate these issues. Parametric calibration methods are preferred for their lower complexity, modeling deviations from ideal responses through perturbation models. These models can be estimated via optimization \cite{appadwedulaManifoldLearningApproach2019, weiCalibrationPhaseShifter2020} or DL-based methods \cite{alrabeiahNeuralNetworksBased2022, iyeDeepLearningBased2022, chenNovelFastFarField2025}. While optimization-based methods can be complex due to numerous unknowns, offline DL approaches struggle with time-varying impairments such as temperature-sensitive phase errors, necessitating periodic recalibration. This limitation highlights the need for online self-calibration frameworks that can adapt to hardware variations in practical systems.

{\bf Contributions and Organization}.
In this study, we propose a DL-based low-overhead beam alignment method for mmWave massive MIMO systems. Our contributions are summarized as follows:
\begin{itemize} 
    \item \emph{Quaternary Search-Based Super-Resolution Beam Alignment:} We analyze the power gains of two intra-layer beams in a DFT codebook and reveal that their ratio exhibits a monotonic trend over the angular interval spanned by the beams. By further characterizing the sensitivity of this power-ratio behavior to noise, we develop a Quaternary Search-based Super-Resolution (QSSR) beam alignment algorithm. The proposed QSSR framework enables low-overhead estimation of the angle of arrival (AoA) and angle of departure (AoD). Instead of selecting the nearest discrete codeword from a predefined codebook, the array directly steers the beam toward the estimated continuous angle, thereby substantially improving beam alignment performance.

    \item \emph{GRU-Based Deep Neural Network for Robust Beam Alignment:}  
    To further improve performance and address the boundary cases of the QSSR algorithm, we exploit additional beam measurements and design a GRU-based recurrent neural network (RNN), referred to as QSSR-Net. This network learns the nonlinear mapping from sequentially collected, normalized power measurements to continuous AoA/AoD estimates. The sequential nature of hierarchical codebook-based beam probing is naturally modeled as a temporal process, enabling the GRU to effectively capture dependencies and extract fine-grained angular information. As a result, QSSR-Net achieves super-resolution beam prediction with low complexity.
    
    \item \emph{Self-Supervised Antenna Calibration for Hardware Impairments:}  
    To mitigate performance degradation caused by practical hardware impairments, we propose a self-supervised antenna calibration method that requires no additional hardware or measurement overhead. By explicitly modeling the impairments, the distorted channels are reconstructed using the outputs of QSSR-Net. The proposed framework, QSSR-Net-Impair, then performs online optimization to update compensation parameters in real time, significantly improving beam alignment under non-ideal hardware conditions.

    \item \emph{Robustness and Generalization in Practical Deployment:}  
    Extensive simulation results demonstrate that both QSSR and QSSR-Net consistently outperform binary search and even surpass exhaustive search at high SNR in both simulated and real-world scenarios (e.g., DeepMIMO \cite{alkhateebDeepMIMOGenericDeep2019}), highlighting their strong generalization capability. Furthermore, the proposed QSSR-Net-Impair framework maintains robust performance under hardware impairments, validating the effectiveness of the self-supervised calibration method.
\end{itemize}

The remainder of this paper is organized as follows. Section~\ref{sec_system_model} introduces the system model and formulates the beam alignment problem. Section~\ref{sec_QSSR} analyzes the power ratio pattern, presents the proposed QSSR algorithm, and further extends it with deep learning techniques. Section~\ref{sec_Impair} describes the self-calibration method for hardware impairments. Simulation results are provided in Section~\ref{sec_simulation_result}, and conclusions are drawn in Section~\ref{sec_conclusion}.

The notations used throughout this paper are as follows. Let $a$, $\mathbf{a}$, and $\mathbf{A}$ denote a scalar, a column vector, and a matrix, respectively. The conjugate, transpose, and conjugate transpose of $\mathbf{A}$ are represented by $\mathbf{A}^{*}$, $\mathbf{A}^T$, and $\mathbf{A}^H$, respectively. The Frobenius norm of $\mathbf{A}$ is denoted by $\|\mathbf{A}\|$.

\section{System Model and Problem Formulation}
\label{sec_system_model}
In this section, we first introduce the system model and then briefly formulate the beam alignment problem. Finally, we discuss the limitations of conventional codebook-based methods, which motivate the super-resolution beam alignment.

\subsection{System Model}

\begin{figure}[tbp!]
    \centering
    \includegraphics[width=0.5\textwidth]{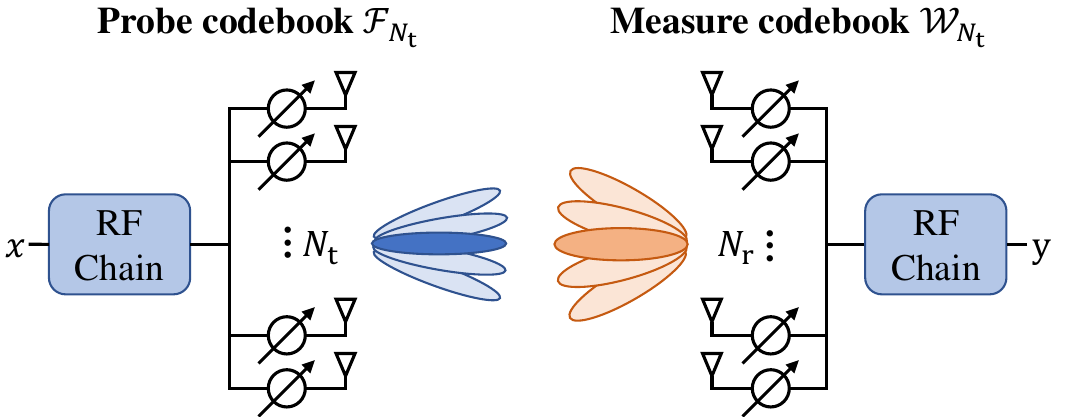}
    \caption{System model of a single-user mmWave massive MIMO system with $N_{\mathrm{t}}$ transmit and $N_{\mathrm{r}}$ receive antennas.}
    \label{fig_system_model}
    \vspace{-10pt}
\end{figure}

We consider a single-user mmWave massive MIMO system in Fig.~\ref{fig_system_model} over a narrowband block-fading channel. The transmitter and receiver are equipped with uniform linear arrays (ULAs) consisting of $N_{\mathrm{t}}$ and $N_{\mathrm{r}}$ antennas, respectively.\footnote{Although a simplified fully-analog ULA-based system is considered for clarity, the proposed method is readily extensible to uniform planar arrays (UPAs), cross-polarized arrays, and hybrid architectures.} The inter-element spacing is assumed to be half the carrier wavelength, i.e., $\lambda/2$. An analog beamforming architecture is adopted, in which each antenna is equipped with a phase shifter, a power amplifier to drive the antenna, and an RF switch for antenna activation \cite{xiaoHierarchicalCodebookDesign2016}. For simplicity, the latter two components are not depicted in the figure. Let $\mathbf{f} \in \mathbb{C}^{N_{\mathrm{t}} \times 1}$ and $\mathbf{w} \in \mathbb{C}^{N_{\mathrm{r}} \times 1}$ denote the transmit and receive beamforming vectors, respectively. Define $\mathcal{I}_{\mathrm{t}} \subseteq \{1, \dots, N_{\mathrm{t}}\}$ and $\mathcal{I}_{\mathrm{r}} \subseteq \{1, \dots, N_{\mathrm{r}}\}$ as the sets of active antennas at the transmitter and receiver. When an RF switch is turned off, the corresponding antenna element is deactivated, resulting in a zero entry in the beamforming vector (e.g., $f_n = 0$ for $n \notin \mathcal{I}_{\mathrm{t}}$). Due to the constant-modulus constraint imposed by phase shifters, the active antenna elements satisfy $|f_n| = \frac{1}{\sqrt{N_{\mathrm{t}}}}$ for $n \in \mathcal{I}_{\mathrm{t}}$ and $|w_m| = \frac{1}{\sqrt{N_{\mathrm{r}}}}$ for $m \in \mathcal{I}_{\mathrm{r}}$. During the data transmission phase, the full arrays are activated, i.e., $\mathcal{I}_{\mathrm{t}} = \{1, \dots, N_{\mathrm{t}}\}$ and $\mathcal{I}_{\mathrm{r}} = \{1, \dots, N_{\mathrm{r}}\}$, to maximize the beamforming gain.

The transmitter sends a unit-power pilot or data symbol $x \in \mathbb{C}$ with $\mathbb{E}[|x|^2] = 1$, precoded by the transmit beamforming vector $\mathbf{f}$. The receiver applies the receive beamforming vector $\mathbf{w}$ to the received signal and obtains  
\begin{equation}
    y = \mathbf{w}^H \mathbf{H} \mathbf{f}\, x + \mathbf{w}^H \mathbf{n},
    \label{equ_signal_model}
\end{equation}
where $\mathbf{H}\in \mathbb{C}^{N_{\mathrm{r}} \times N_{\mathrm{t}}}$ denotes the channel between the transmitter and receiver, and $\mathbf{n} \in \mathbb{R}^{N_{\mathrm{r}}\times 1} \sim \mathcal{CN}(\mathbf{0}, \sigma^2 \mathbf{I}_{N_{\mathrm{r}}})$ represents additive white Gaussian noise with variance $\sigma^2$. 

\subsection{Problem Formulation}

To facilitate the following formulation, we model the mmWave channel using the Saleh-Valenzuela (S-V) model, which captures the sparse scattering nature of mmWave propagation. The channel matrix is expressed as
\begin{equation} 
    \mathbf{H} = \sum_{l=1}^{L} \alpha_l \, \mathbf{a}(N_{\mathrm{r}}, \theta_l) \, \mathbf{a}^H(N_{\mathrm{t}}, \phi_l),
    \label{equ_SV_model}
\end{equation}
where $L$ is the number of resolvable propagation paths, $\alpha_l$ denotes the complex gain of the $l$-th path, $\theta_l \in [-1, 1]$ is the normalized angle of arrival (AoA) at the receiver, and $\phi_l \in [-1, 1]$ is the normalized angle of departure (AoD) at the transmitter. Without loss of generality, we assume $|\alpha_1| > \cdots > |\alpha_L|$. The array steering vector for an $N_{\mathrm{t}}$-element ULA with half-wavelength spacing is defined as
\begin{equation}
    \mathbf{a}(N_{\mathrm{t}}, \vartheta) = \frac{1}{\sqrt{N_{\mathrm{t}}}}
    \left[ 1, \ e^{j\pi\vartheta}, \ \dots, \ e^{j(N_{\mathrm{t}}-1)\pi\vartheta} \right]^T.
    \label{equ_steering_vector}
\end{equation}

The goal of beam alignment is to determine the transmit and receive beamforming vectors that maximize the received signal power, leading to the optimization problem
\vspace{-3pt}
\begin{align} 
    \max_{\mathbf{f}, \mathbf{w}} \quad & \big|\mathbf{w}^H \mathbf{H} \mathbf{f}\big|^2 \notag
    \\
    \text{s.t.} \quad & |f_n| = \frac{1}{\sqrt{N_{\mathrm{t}}}}, 
    \ |w_n| = \frac{1}{\sqrt{N_{\mathrm{r}}}}, \ \forall n.
    \label{equ_opt_problem}
\end{align}
\vspace{-10pt}

To efficiently address this problem, beam search is widely adopted in practice, wherein the beamforming vectors are selected from predefined codebooks. A common choice is the discrete Fourier transform (DFT) codebook, which uniformly samples the angular domain as
\vspace{-5pt}
\begin{multline} 
    \mathcal{F}_{N_{\mathrm{t}}} \!=\! \Big\{ \mathbf{a}\! \left(N_{\mathrm{t}}, \vartheta_{N_\mathrm{t},p}\right) \;\Big|  \; 
    \vartheta_{N_\mathrm{t},p} = -1 + \tfrac{2p-1}{N_{\mathrm{t}}}, \\
    p = 1, \ldots, N_{\mathrm{t}} \Big\},
\end{multline}
where each codeword corresponds to a beam steered toward a quantized angle. The $p$-th codeword in $\mathcal{F}_{N_{\mathrm{t}}}$, denoted by $\boldsymbol{f}_{N_{\mathrm{t}},p}$, has the pointing direction $\vartheta_{N_{\mathrm{t}},p} = -1 + \frac{2p-1}{N_{\mathrm{t}}}$. The receive codebook $\mathcal{W}_{N_{\mathrm{r}}}$ is defined in the same manner.

Under this setting, the solution of (\ref{equ_opt_problem}) can be obtained by searching over all $N_{\mathrm{t}} N_{\mathrm{r}}$ beam pairs and selecting the one that maximizes the received power. Although this guarantees satisfactory performance, the resulting measurement overhead is prohibitive for large arrays. To reduce complexity, hierarchical search strategies such as binary search \cite{xiaoHierarchicalCodebookDesign2016} have been proposed, where the beam is progressively refined, starting from wide beams in a coarse codebook (e.g., $\mathcal{F}_{4}$) and narrowing to finer beams in deeper layers (e.g., $\mathcal{F}_{8}$). Here, $\mathcal{F}_{2^{l}}$ denotes the $l$-th layer codebook constructed using the first $2^{l}$ antennas. The refinement is guided by previous measurements, since each wide beam encompasses multiple narrower beams, and the optimal finer beam typically lies within the coverage of the selected coarse beam. This progressive narrowing significantly reduces the number of measurements compared to exhaustive search, making hierarchical beam alignment more practical for large-scale antenna arrays.

\begin{figure}[tbp!]
    \centering
    \includegraphics[width=0.4\textwidth]{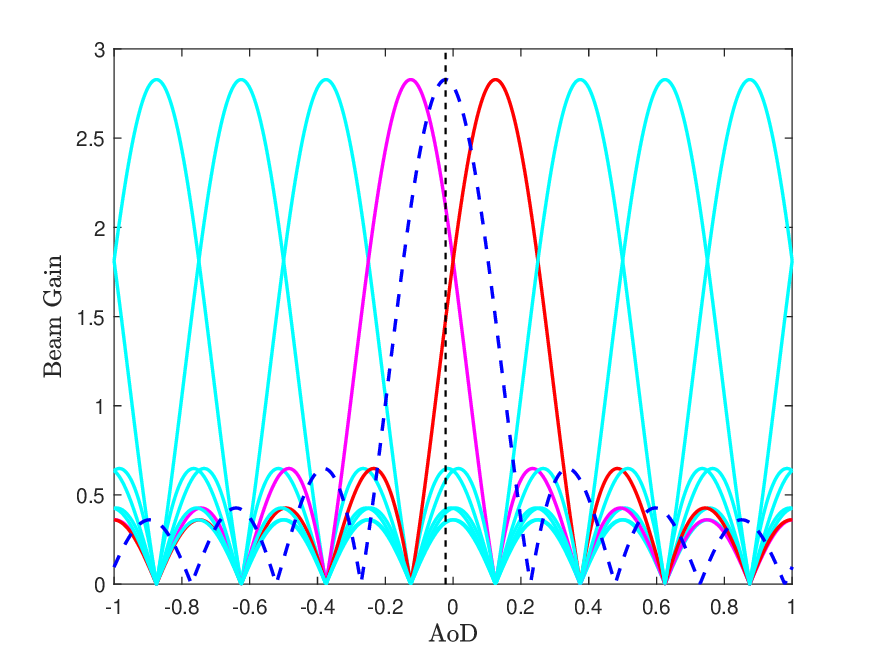}
    \vspace{-10pt}
    \caption{Illustration of the DFT codebook with $N_{\mathrm{t}} = 8$.}
    \label{fig_DFT_Codebook_Shape}
    \vspace{-15pt}
\end{figure}

\begin{figure}[tbp!]
    \centering
    \includegraphics[width=0.4\textwidth]{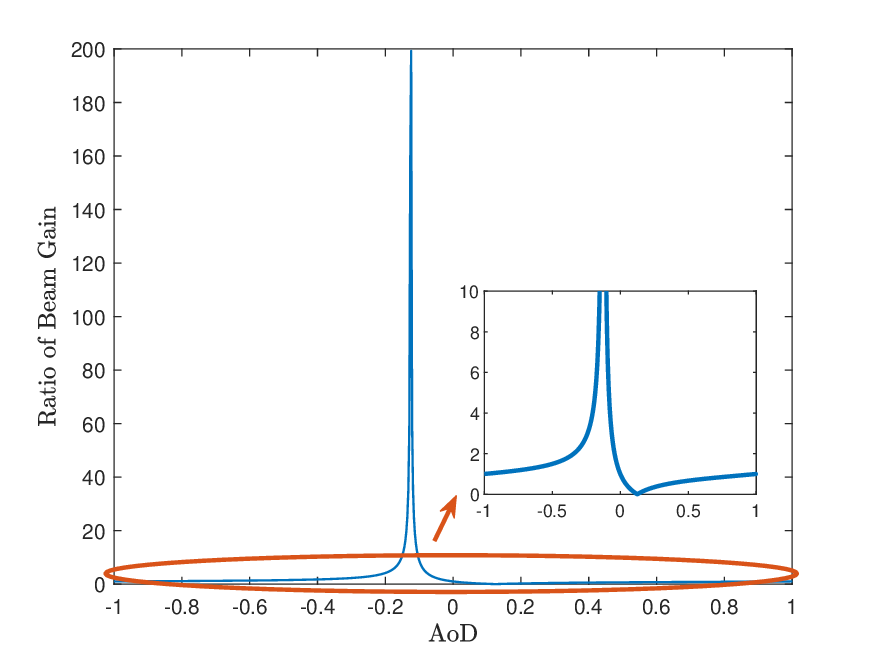}
    \vspace{-10pt}
    \caption{Ratio of beam gains for the fourth and fifth codewords, $\boldsymbol{f}_{8,4}$ and $\boldsymbol{f}_{8,5}$, in the DFT codebook with $N_{\mathrm{t}} = 8$.}
    \label{fig_DFT_Codebook_Ratio}
    \vspace{-16pt}
\end{figure}

However, the angular quantization inherent in finite codebooks imposes a resolution limit. Even with large arrays and correspondingly larger codebooks, the selected beam pair may still suffer from non-negligible pointing errors. Fig.~\ref{fig_DFT_Codebook_Shape} illustrates the DFT codebook generated by $N_{\mathrm{t}} = 8$ antennas. Although the eight beams uniformly cover the angular domain, the beamforming gain exhibits a shell-like pattern across space. This phenomenon, known as the \emph{shelling effect}, arises because the actual AoD rarely coincides with a codebook beam center. As the angle approaches the boundary between adjacent beams, the array response deviates from its peak, leading to gain degradation. For example, if the actual AoD corresponds to the black dashed line, beam search with the DFT codebook will select the magenta beam, which incurs a noticeable gain loss compared to the virtual beam represented by the blue dashed line. Importantly, this effect persists even when the codebook size increases, revealing the inherent quantization barrier of conventional approaches. These limitations motivate the development of \emph{super-resolution} beam alignment methods. Once the AoA and AoD of the dominant path, denoted by $\theta_1$ and $\phi_1$, are estimated, the beams can be steered accordingly as $\mathbf{f} = \mathbf{a}(N_{\mathrm{t}}, \phi_1)$ and $\mathbf{w} = \mathbf{a}(N_{\mathrm{r}}, \theta_1)$, instead of being selected from a predefined codebook. However, accurately estimating continuous angles from received signals remains challenging due to high training overhead and computational complexity.

\vspace{-5pt}
\section{Quaternary Search-Based Super-Resolution Beam Alignment}
\label{sec_QSSR}
In this section, we propose a super-resolution beam alignment method to overcome the quantization limitation of codebook-based schemes. We first analyze the beam gain behavior and establish its monotonic property. Leveraging this property, we design a quaternary search strategy to efficiently localize the angular range of the optimal beam direction. The established monotonicity is then utilized to estimate the target beam direction. Finally, we incorporate DL techniques to further enhance the angle estimation accuracy, thereby improving the overall beam alignment performance.

\subsection{Beam Gain Behavior of Two Beams in the Same Layer}

From Fig.~\ref{fig_DFT_Codebook_Shape}, we observe that each beam points in a specific direction, and between two adjacent beams, the gain of one increases monotonically while that of the other decreases monotonically. Motivated by this, we plot the ratio of beam gains for the fourth and fifth codewords in the DFT codebook (the magenta and red beams in Fig.~\ref{fig_DFT_Codebook_Shape}), as shown in Fig.~\ref{fig_DFT_Codebook_Ratio}. Here, the 4th beam points at $\vartheta_{8,4} = -\tfrac{1}{8}$ and the 5th beam points at $\vartheta_{8,5} = \tfrac{1}{8}$. The figure clearly shows that the gain ratio of the two codewords varies monotonically within specific intervals. In particular, the ratio of the 4th and 5th codewords decreases monotonically over $[\vartheta_{8,4}, \vartheta_{8,5}]$ and increases monotonically over $[-1, \vartheta_{8,4}]$ and $[\vartheta_{8,5}, 1]$. 

Without loss of generality, we focus on transmitter-side beam search and provide a theoretical justification for this property. We consider a line-of-sight (LoS)-dominant scenario, where the receiver employs a quasi-omnidirectional pattern (or a fixed beamformer) to probe the channel. Let $\phi$ denote the AoD. The effective beamforming gain associated with the $p$-th transmit codeword can be expressed as
\begin{align}
        G_p(\phi)
        &= \big| \boldsymbol{f}_{N_{\mathrm{t}}, p}^H \mathbf{a}(N_{\mathrm{t}},\phi)\big|^2
        = \frac{1}{N_{\mathrm{t}}^2} \left| \sum_{n=0}^{N_{\mathrm{t}}-1} e^{j n \pi (\phi-\vartheta_{N_{\mathrm{t}}, p})} \right|^2 \notag \\
        &= \frac{1}{N_{\mathrm{t}}^2} \left| \frac{1-e^{jN_{\mathrm{t}} \pi (\phi-\vartheta_{N_{\mathrm{t}}, p})}}{1-e^{j\pi(\phi-\vartheta_{N_{\mathrm{t}}, p})}} \right|^2 \notag \\
        & = \frac{1}{N_{\mathrm{t}}^2}\left(\frac{\sin\!\big(\tfrac{N_{\mathrm{t}}\pi(\phi-\vartheta_{N_{\mathrm{t}}, p})}{2}\big)}{\sin\!\big(\tfrac{\pi(\phi-\vartheta_{N_{\mathrm{t}}, p})}{2}\big)}\right)^{\!2}. \label{equ_beam_gain_dirichlet}
\end{align} 
The complex path gain is omitted, as it naturally cancels out in the subsequent ratio analysis. Considering two codewords indexed by $p$ and $q$, we define the gain ratio as
\begin{align}
    \rho_{p,q}(\phi) &\triangleq \frac{G_p(\phi)}{G_q(\phi)} \notag \\
    & = \left(\frac{\sin\!\big(\tfrac{N_{\mathrm{t}}\pi(\phi-\vartheta_{N_{\mathrm{t}}, p})}{2}\big)}{\sin\!\big(\tfrac{\pi(\phi-\vartheta_{N_{\mathrm{t}}, p})}{2}\big)}
    \cdot
    \frac{\sin\!\big(\tfrac{\pi(\phi-\vartheta_{N_{\mathrm{t}}, q})}{2}\big)}{\sin\!\big(\tfrac{N_{\mathrm{t}}\pi(\phi-\vartheta_{N_{\mathrm{t}}, q})}{2}\big)}
    \right)^{\!2}.
    \label{equ_ratio_exact}
\end{align}
For any two codewords indexed by $p$ and $q$ in the same layer of the DFT codebook, their pointing directions satisfy $\vartheta_{N_{\mathrm{t},} q} = \vartheta_{N_{\mathrm{t},} p} + \frac{2k}{N_{\mathrm{t}}}$, where $k = q - p$ represents the integer index separation. Substituting this into \eqref{equ_ratio_exact} yields
\begin{align}
    \rho_{p,q}(\phi) 
    &= \left(\frac{\sin\!\big(\tfrac{N_{\mathrm{t}}\pi(\phi-\vartheta_{N_{\mathrm{t},} p})}{2}\big)}{\sin\!\big(\tfrac{\pi(\phi-\vartheta_{N_{\mathrm{t},} p})}{2}\big)}
        \cdot
        \frac{\sin\!\big(\tfrac{\pi(\phi-\vartheta_{N_{\mathrm{t},} p})}{2} - \frac{k\pi}{N_{\mathrm{t}}}\big)}{\sin\!\big(\tfrac{N_{\mathrm{t}}\pi(\phi-\vartheta_{N_{\mathrm{t},} p})}{2} - k\pi \big)}
        \right)^{\!2} \nonumber \\
    &= \left(\frac{\sin\!\big(\tfrac{\pi(\phi-\vartheta_{N_{\mathrm{t},} p})}{2} - \frac{k\pi}{N_{\mathrm{t}}}\big)}{\sin\!\big(\tfrac{\pi(\phi-\vartheta_{N_{\mathrm{t},} p})}{2}\big)} \right)^{\!2}.
    \label{equ_ratio_simplified}
\end{align}

The monotonicity of $\rho_{p,q}(\phi)$ can be analyzed by differentiating \eqref{equ_ratio_simplified} as 
\begin{equation}
\label{equ_derivative}
    \frac{d\rho_{p,q}(\phi)}{d\phi} = \frac{\pi \sin(\frac{\pi k}{N_{\mathrm{t}}})\sin\big(\frac{\pi(\phi-\vartheta_{N_{\mathrm{t},} p})}{2} - \frac{\pi k}{N_{\mathrm{t}}}\big)}{\sin^3\!\big(\frac{\pi(\phi-\vartheta_{N_{\mathrm{t},} p})}{2}\big)}.
\end{equation}
We analyze the sign of $\tfrac{d\rho}{d\phi}$ in two intervals:
\begin{itemize}
    \item \textbf{Interval $(\vartheta_{N_{\mathrm{t}},p}, \vartheta_{N_{\mathrm{t}},q})$:}
    Here $0 < \phi - \vartheta_{N_{\mathrm{t}},p} < \tfrac{2k}{N_{\mathrm{t}}}$.
    The denominator $\sin^{3}(\cdot)$ is positive, while
    $\sin\!\left(\tfrac{\pi(\phi-\vartheta_{N_{\mathrm{t}},p})}{2} - \tfrac{\pi k}{N_{\mathrm{t}}}\right)$ is negative.
    Therefore, $\tfrac{d\rho}{d\phi} < 0$, and $\rho_{p,q}(\phi)$ decreases monotonically.
    As an illustrative example, consider $N_{\mathrm{t}} = 8$, $p = 4$, $q = 5$, and $k = 1$.
    When $\phi \in [\vartheta_{8,4}, \vartheta_{8,5}]$, we have $0 < \phi - \vartheta_{8,4} < \tfrac{2}{N_{\mathrm{t}}}$,
    and the gain ratio $\rho_{p,q}(\phi)$ decreases monotonically, consistent with the trend in Fig.~\ref{fig_DFT_Codebook_Ratio}.
    
    \item \textbf{Interval $(\vartheta_{N_{\mathrm{t}},q}, \vartheta_{N_{\mathrm{t}},p} + 2)$:}  
    Here $\tfrac{2k}{N_{\mathrm{t}}} < \phi - \vartheta_{N_{\mathrm{t}},p} < 2$.  
    Both the denominator and the shifted sine term are positive, and $\sin\!\left(\tfrac{\pi k}{N_{\mathrm{t}}}\right) > 0$,  
    yielding $\tfrac{d\rho}{d\phi} > 0$. Hence, $\rho_{p,q}(\phi)$ increases monotonically.  
    Again, with $N_{\mathrm{t}} = 8$, $p = 4$, $q = 5$, and $k = 1$,  
    we find that when $\phi \in [\vartheta_{8,5}, \vartheta_{8,4} + 2]$,  
    the gain ratio $\rho_{p,q}(\phi)$ increases monotonically.  
    Since $\rho_{p,q}(\phi)$ is periodic in $\phi$ with period 2,  
    this interval can equivalently be expressed as $[-1, \vartheta_{8,4}] \cup [\vartheta_{8,5}, 1]$,  
    which aligns with the observations in Fig.~\ref{fig_DFT_Codebook_Ratio}.
\end{itemize}
This confirms that the monotonic gain ratio property holds for any two beams in the same DFT layer. Furthermore, noting that $\rho_{p,q}(\vartheta_{N_{\mathrm{t}}, p}) = +\infty$ and $\rho_{p,q}(\vartheta_{N_{\mathrm{t}}, q}) = 0$, it follows that for any power ratio $\gamma = \rho_{p,q}(\phi)$, there exists a unique angle $\phi \in [\vartheta_{N_{\mathrm{t}}, p}, \vartheta_{N_{\mathrm{t}}, q}]$ satisfying $\phi = \rho_{p,q}^{-1}(\gamma)$. In other words, the mapping between the gain ratio and the AoD within the interval spanned by the two beams is bijective.

Beyond the uniqueness of the mapping, the derivative in \eqref{equ_derivative} provides key insights into the sensitivity of angle estimation to noise. Let $\Delta \rho$ denote the perturbation in the power ratio induced by noise and multipath components. Using a first-order Taylor expansion, the resulting estimation error $\Delta \phi$ can be approximated as
\begin{equation}
    \Delta \phi \approx \frac{d \rho_{p,q}^{-1}}{d \rho}  \Delta \rho
    = \frac{\Delta \rho}{\frac{d\rho_{p,q}(\phi)}{d\phi}}.
    \label{equ_error_sensitivity_theory}
\end{equation}
This expression indicates that the estimation accuracy depends on two factors: the magnitude of the perturbation $\Delta \rho$ and the slope of the ratio function, $\left| \frac{d\rho_{p,q}(\phi)}{d\phi} \right|$. First, the perturbation $\Delta \rho$ is inversely related to the received signal power, such that beams with low gain amplify the effect of noise. Second, the estimation error decreases with increasing gradient magnitude, implying that the high-slope regions between adjacent beams provide maximum robustness against noise. Consequently, although any beam pair in principle defines a monotonic mapping, only the adjacent beams enclosing the target angle simultaneously maximize the received power (minimizing $\Delta \rho$) and the gradient (maximizing $\left| \frac{d\rho_{p,q}}{d\phi} \right|$). This theoretical insight underpins the beam selection strategy in the proposed algorithm, which dynamically chooses anchor and auxiliary beams to ensure estimation occurs within high-gradient regions.

\subsection{Proposed QSSR Algorithm}

Based on the derived bijection and sensitivity properties, intra-layer power measurements can be exploited for reliable and low-overhead angle estimation. The practical selection rule is simple yet effective: choose the beam with the largest received power as the \emph{anchor beam} and select its strongest neighboring beam as the \emph{auxiliary beam}. By calculating the power ratio of these two beams, the corresponding angle can be readily estimated. Motivated by this principle, we propose the QSSR algorithm to implement this strategy within a hierarchical beam alignment framework. The detailed procedure is elaborated as follows.

To avoid the prohibitive overhead associated with exhaustive scanning, we adopt a hierarchical search framework. However, a direct application of the conventional \emph{binary search} is ill-suited for ratio-based super-resolution estimation due to an intrinsic geometric limitation. Consider the refinement from a coarse layer with $N_{\mathrm{t}}/2$ activated antennas to the finest layer with $N_{\mathrm{t}}$ activated antennas. If the coarse search selects the $p$-th codeword, the true angle is confined to a sector $\mathcal{R}_{\mathrm{total}}= \left[-1 + \frac{4p-4}{N_{\mathrm{t}}},\, -1 + \frac{4p}{N_{\mathrm{t}}}\right],$ whose angular span is $4/N_{\mathrm{t}}$. In the subsequent binary refinement, two fine beams indexed by $2p-1$ and $2p$ are probed, with steering directions $\vartheta_{N_{\mathrm{t}},2p-1} = -1 + \frac{4p-3}{N_{\mathrm{t}}}, \vartheta_{N_{\mathrm{t}},2p} = -1 + \frac{4p-1}{N_{\mathrm{t}}}$. As established in the previous section, accurate ratio-based angle estimation is guaranteed only when the true angle lies within the region enclosed by the peaks of these two beams, namely, $\mathcal{R}_{\mathrm{valid}}= \big[\vartheta_{N_{\mathrm{t}},2p-1},\, \vartheta_{N_{\mathrm{t}},2p}\big]$. The length of this valid interval is $2/N_{\mathrm{t}}$, which covers only $50\%$ of the refined sector $\mathcal{R}_{\mathrm{total}}$. If the true angle falls into the remaining half of $\mathcal{R}_{\mathrm{total}}$, corresponding to the outer edges of the beam pair, the ratio-based estimation becomes ambiguous and inaccurate. 

To mitigate this limitation while preserving logarithmic training overhead, we adopt a \emph{quaternary search} strategy. Instead of probing two beams per refinement step, the proposed scheme simultaneously probes four adjacent fine beams indexed from $p$ to $p+3$. Notably, since $4 \log_{4} N_{\mathrm{t}} = 2 \log_{2} N_{\mathrm{t}}$, the quaternary search incurs the same overall measurement overhead as the conventional binary search, while offering a substantially improved geometric coverage. Specifically, the union of the three adjacent valid intervals formed by consecutive beam pairs, $\mathcal{R}_{\mathrm{valid}}^{\mathrm{quat}} = \bigcup_{i=0}^{2} \left[ \vartheta_{N_{\mathrm{t}},\,p+i},\, \vartheta_{N_{\mathrm{t}},\,p+i+1} \right]$, covers approximately $75\%$ of the scanned angular sector. This significantly reduces the probability that the true angle lies in regions where ratio-based estimation becomes ambiguous or inaccurate. Moreover, the availability of four candidate beams enables the algorithm to dynamically select the optimal anchor and auxiliary beam pair that encloses the true angle. As a result, the angle estimation is consistently performed within a high-slope region of the gain ratio, thereby minimizing sensitivity to noise and multipath interference, in accordance with the theoretical sensitivity analysis presented earlier.

Algorithm~\ref{alg_QSSR} summarizes the overall procedure. Beam alignment is performed sequentially at the transmitter and then at the receiver. During the transmitter-side search, the receiver adopts a quasi-omnidirectional beam by activating a single antenna, i.e., $\mathbf{w} = [1/\sqrt{N_{\mathrm{r}}},0,\ldots,0]^T$. The transmitter then performs a quaternary search starting from the coarsest codebook $\mathcal{F}_4$, and recursively selects the most promising quartet of beams to probe in the next finer layer (e.g., $\mathcal{F}_{16}$). At the final layer, the anchor beam is identified as the beam that yields the maximum received power. An auxiliary beam is subsequently selected from its neighboring beams. Crucially, if the anchor beam lies at the boundary of the quartet, such that the optimal neighboring beam may not have been probed, the algorithm defaults to selecting the remaining available neighbor as the auxiliary beam. This situation corresponds to the $25\%$ ambiguity region identified in the previous analysis. Although incorrect auxiliary beam selection produces large errors, simulations show that the proposed scheme attains satisfactory performance in practice, and these errors are addressed by the DL-empowered refinement described later.

Once the anchor and auxiliary beams are identified, their measured power ratio is computed and mapped to an AoD estimate via the bijection established over the corresponding interval. Since the power ratio function in \eqref{equ_ratio_exact} involves transcendental trigonometric expressions, a closed-form analytical inverse is generally unavailable. In practice, we therefore adopt a numerical inversion strategy. Specifically, the bisection algorithm is employed during simulations to efficiently recover the angle estimate. After obtaining the AoD estimate $\hat{\phi}$, the transmitter sets $\mathbf{f}=\mathbf{a}(N_{\mathrm{t}},\hat{\phi})$ and the receiver-side quaternary search proceeds analogously to produce the AoA estimate $\hat{\theta}$.

\begin{algorithm}[tbp!]
    \SetKwInOut{Input}{Input}\SetKwInOut{Output}{Output}
    \let\oldnl\nl
    \newcommand{\nonl}{\renewcommand{\nl}{\let\nl\oldnl}}
    \Output{Estimated transmit beam \( \mathbf{f} \), receive beam \( \mathbf{w} \)}  
    \vspace{1pt}
    \nonl \textbf{Transmitter Side Search:}\\
    Initialize \( \mathbf{w} = [1/{\sqrt{N_{\mathrm{r}}}}, 0, \dots, 0]^T \), \( i_{\mathrm{t}} = 1 \)\;
    \For{$l = 1 : \lfloor \log_{4} N_{\mathrm{t}}\rfloor$}{
         Scan the \( 4i_{\mathrm{t}}-3 \)-th to \( 4i_{\mathrm{t}} \)-th codewords in the codebook \( \mathcal{F}_{4^l} \)\;
        Record received power \( \mathbf{p}_{\mathrm{t}} \in \mathbb{R}^4 \)\;
        \( i_{\mathrm{t}} = \arg\max_{i} \mathbf{p}_{\mathrm{t}}[i] \)\;
    }
    Identify anchor index \( i_{\mathrm{t, anchor}} = \arg\max_{i} \mathbf{p}_{\mathrm{t}}[i] \)\;
    Determine adjacent auxiliary beam\;
    Calculate the received power ratio and inverse map the estimated AoD $\hat{\phi}$\;
    Set \( \mathbf{f} = \mathbf{a}(N_{\mathrm{t}}, \hat{\phi}) \).  
    
    \vspace{1pt}
    \nonl \textbf{Receiver Side Search:}\\
    Initialize \( i_{\mathrm{r}} = 1 \)\;
    \For{$l = 1 : \lfloor \log_{4} N_{\mathrm{r}}\rfloor$}{
        Scan the \( 4i_{\mathrm{r}}-3 \)-th to \( 4i_{\mathrm{r}} \)-th codewords in the codebook \( \mathcal{W}_{4^l} \)\;
        Record received power \( \mathbf{p}_{\mathrm{r}} \in \mathbb{R}^4 \)\;
        \( i_{\mathrm{r}} = \arg\max_{i} \mathbf{p}_{\mathrm{r}}[i] \)\; 
    }
    Identify anchor beam \( i_{\mathrm{r, anchor}} = \arg\max_{i} \mathbf{p}_{\mathrm{r}}[i] \)\;
    Determine adjacent auxiliary beam\;
    Calculate the received power ratio and inverse map the estimated AoA $\hat{\theta}$.\;
    Set \( \mathbf{w} = \mathbf{a}(N_{\mathrm{r}}, \hat{\theta}) \)\;

    \caption{Proposed QSSR Beam Alignment Algorithm}
    \label{alg_QSSR}
\end{algorithm}
\begin{figure}[tb]
    \centering
    \includegraphics[width=0.48\textwidth]{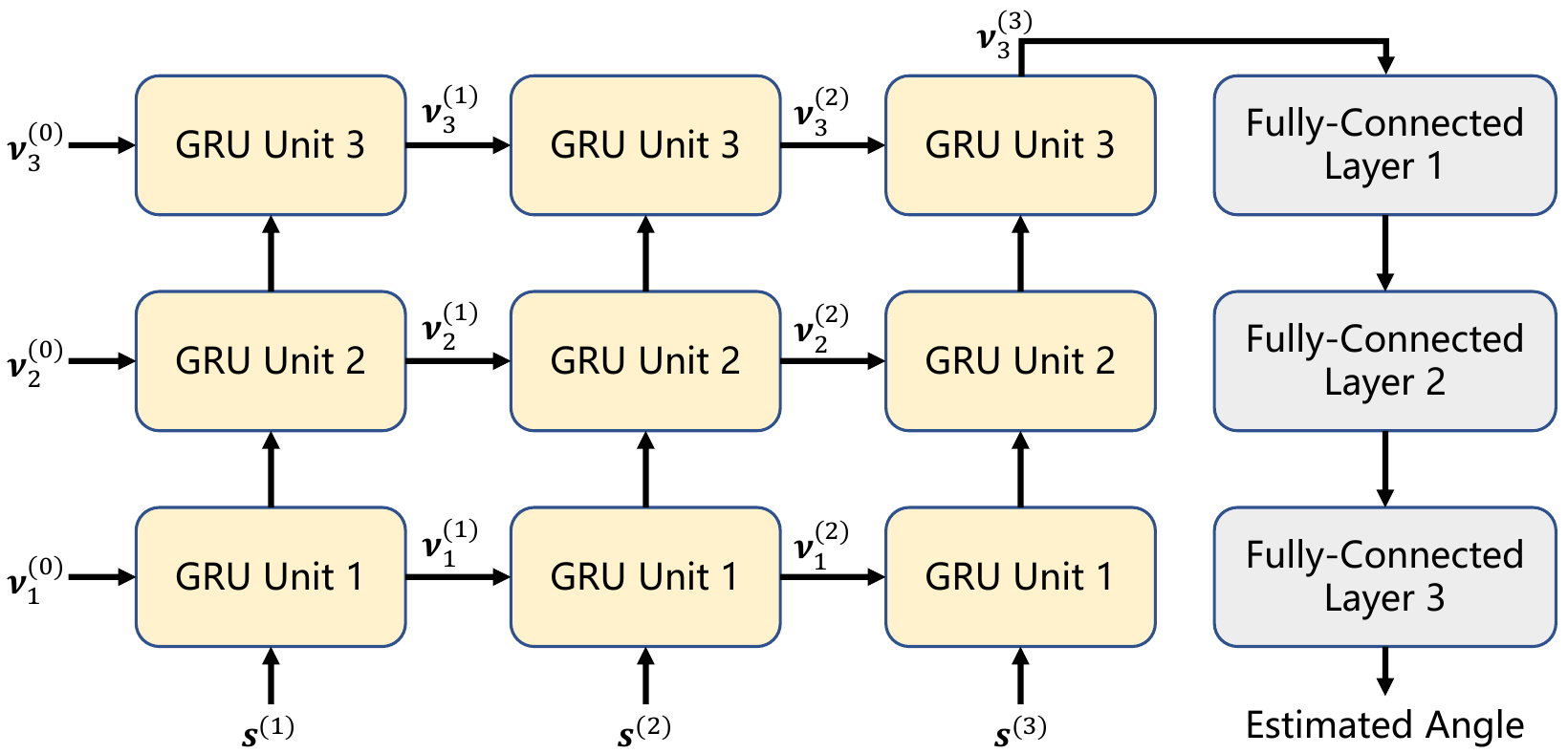}
    \vspace{-5pt}
    \caption{Network structure of the proposed QSSR-Net. }
    \label{fig_Network_Structure}
    \vspace{-15pt}
\end{figure}

\vspace{-5pt}
\subsection{DL-Empowered QSSR-Net}

In practice, arbitrary beam pairs within the same layer also exhibit monotonic behavior, and intermediate measurements obtained from earlier layers can provide additional informative cues for angle inference. However, the original QSSR method utilizes only a limited portion of this information, leading to underexploitation of available estimation evidence. To jointly utilize power ratios from all measured beams, a mapping from multi-layer beam measurements to the underlying angular parameter is required. While a deterministic monotonic mapping can be analytically derived for a single adjacent beam pair, it becomes analytically intractable to construct a globally monotonic and closed-form mapping when multiple gain ratios are jointly incorporated. This observation motivates the introduction of DL, which is well suited for approximating highly nonlinear mappings. By leveraging the strong representation capability of DL, QSSR-Net can capture implicit dependencies among multi-layer beam measurements and effectively fuse information from multiple gain ratios, thereby achieving improved angular estimation accuracy compared with the analytical QSSR approach.

The design of QSSR-Net is explicitly motivated by the structure of the QSSR algorithm. The super-resolution capability fundamentally originates from the power ratio between adjacent beams, which implies that the network input should focus on power ratios rather than raw received powers. While the power ratio captures the relative angular offset within a local interval, recovering the absolute AoA and AoD requires anchoring this offset to the corresponding beam steering directions. Therefore, the beam pointing directions are also included in the input feature set. This feature design ensures that the network input is fully aligned with the variables governing the analytical solution. Furthermore, since the codebook is scanned sequentially across multiple hierarchical layers, the resulting beam measurements naturally form a temporal sequence. This inherent sequential structure motivates the adoption of a RNN. To balance estimation accuracy and computational efficiency, the gated recurrent unit (GRU) is employed. Compared with standard RNNs, GRUs provide improved capability in modeling sequential dependencies, while requiring fewer parameters and lower complexity than long short-term memory networks, making them well suited for low-latency beam alignment applications.

We introduce the input representation and network structure of QSSR-Net. In quaternary search, four beams are measured at each layer, and the received signal power of the $l$-th layer is recorded as ${\mathbf{p}_{\mathrm{t}}^{(l)} \in \mathbb{R}^{4 \times 1}}$. Following the QSSR algorithm, the normalized power vector $\mathbf{p}_{\mathrm{t}}^{(l)}/\max(\mathbf{p}_{\mathrm{t}}^{(l)})$ is used as input. The corresponding pointing directions $\{\bar{\vartheta}_{1}^{(l)}, \bar{\vartheta}_{2}^{(l)}, \bar{\vartheta}_{3}^{(l)}, \bar{\vartheta}_{4}^{(l)}\}$ are also incorporated, forming the input vector
\[
\mathbf{s}^{(l)} = \left\{\bar{\vartheta}_{1}^{(l)}, \bar{\vartheta}_{2}^{(l)}, \bar{\vartheta}_{3}^{(l)}, \bar{\vartheta}_{4}^{(l)}, \mathbf{p}_{\mathrm{t}}^{(l)}/\max(\mathbf{p}_{\mathrm{t}}^{(l)}) \right\} \in \mathbb{R}^{8 \times 1}.
\]

The sequential inputs $\mathbf{s}^{(l)}$ collected across layers are fed into a three-layer GRU network, as shown in Fig.~\ref{fig_Network_Structure}. The hidden state of the $k$-th GRU layer at the $l$-th step is denoted by $\boldsymbol{v}_k^{(l)}$, with initial states $\boldsymbol{v}_1^{(0)}=\boldsymbol{v}_2^{(0)}=\boldsymbol{v}_3^{(0)}=\mathbf{0}$. After the final codebook layer is scanned, the extracted features are passed through three fully connected layers to produce the estimated angle. A ReLU activation function is applied to the intermediate layers to enhance nonlinearity. No activation function is employed at the output layer, allowing the network to freely predict continuous angles in the region $[-1,1]$, thereby enabling correction of potential selection errors introduced during the quaternary search. In practice, any estimate that falls outside this region can be straightforwardly projected back by exploiting the periodicity of the angular domain.

The QSSR-Net functions as an inference engine that substitutes Steps~7--9 and 17--19 in Algorithm~\ref{alg_QSSR}. It observes the entire beam search process in an offline manner, rather than modifying the beam selection path in real time. Since the ultimate objective of beam alignment is to maximize the received signal power, the loss function is defined as the objective in (\ref{equ_opt_problem}). The transmitter-side and receiver-side networks are jointly trained offline using a dataset of simulated channel realizations generated according to the S-V channel model. Once trained, the model can be directly deployed in practical systems and demonstrates strong robustness across diverse propagation environments, as corroborated by the simulation results. Owing to its model-driven design, the proposed QSSR-Net maintains the low beam measurement overhead of the original QSSR framework. In addition, it effectively mitigates the ambiguity caused by auxiliary beam selection, as the network jointly leverages the entire input power profile to produce a refined beam estimate.

\section{Antenna Calibration for Hardware Impairments}
\label{sec_Impair}
Although super-resolution methods can achieve advanced performance, the monotonicity of the power-ratio function critically depends on an ideal DFT codebook. In practical systems, antenna arrays are inevitably affected by hardware impairments, which distort the beam patterns and violate the ideal bijection property. Under such conditions, both the proposed QSSR algorithm and the enhanced QSSR-Net exhibit negligible performance improvement over conventional methods. Consequently, online and periodic self-calibration becomes necessary. In this section, we first introduce different types of hardware impairments, and then develop self-calibration methods to mitigate the performance degradation.

\subsection{Types of Hardware Impairments}

The practical mmWave systems inevitably suffer from hardware imperfections. These impairments can be broadly categorized into long-term and short-term types. The \emph{long-term impairments}, such as mutual coupling between antennas, antenna gain errors, geometric errors of antenna elements, and phase errors caused by circuit asymmetry, originate from element design limitations and physical constraints, and are independent of the transmitted signal. Hence, antenna calibration is primarily designed to compensate for these impairments. In contrast, the \emph{short-term impairments}, including phase noise due to circuit jitter and quantization errors from low-resolution ADC/DACs, vary with the transmitted signals and can be effectively modeled as additional noise.

Although long-term impairments vary slowly, they are not strictly constant. For example, the phase errors caused by circuit asymmetry are temperature-sensitive and may change under different operating conditions. This characteristic limits the effectiveness of traditional antenna calibration methods that assume static impairments. To address this issue, we propose an \emph{online antenna calibration} method that dynamically adapts to varying conditions and is naturally integrated with the proposed super-resolution beam alignment framework.

To simplify the analysis, we focus on antenna calibration for \emph{position errors} and \emph{phase errors}, and examine their impact on the array steering vector.
\begin{itemize}
    \item \textbf{Position Errors:}  
    In the ideal case, the inter-element spacing is assumed to be exactly $d = \lambda/2$. However, mechanical imperfections or thermal expansion may cause deviations from the ideal spacing. Even small deviations accumulate across large arrays. The array response with position errors can be expressed as
    \begin{equation} 
        \mathbf{a}_{\mathrm{pos}}(N_{\mathrm{t}}, \vartheta) = \mathbf{D}(\vartheta)\,\mathbf{a}(N_{\mathrm{t}},\vartheta),
    \vspace{-5pt}
    \end{equation}
    where
    \begin{equation}
        \vspace{-5pt}
        \mathbf{D}(\vartheta) = \operatorname{diag}\!\left(\bigl[1,\; e^{j\frac{2\pi}{\lambda}\delta_{d_2}\vartheta},\; \ldots,\; e^{j\frac{2\pi}{\lambda}\delta_{d_N}\vartheta}\bigr]\right).
    \end{equation}
    Here, $\delta_{d_n}$ denotes the displacement error of the $n$-th element relative to its ideal position. It can be modeled as a Gaussian distribution, $\delta_{d_n} \sim \mathcal{N_{\mathrm{t}}}(0,\sigma_\mathsf{d}^2)$, or as a truncated Gaussian, $\delta_{d_n} \sim \widetilde{\mathcal{N_{\mathrm{t}}}}\!\left(0,\sigma_\mathsf{d}^2;-\tfrac{1}{4}\lambda, \tfrac{1}{4}\lambda\right)$, to enforce physical bounds.
    
    \item \textbf{Phase Errors:}  
    RF circuits, including cables and connectors between RF chains and antenna ports, inevitably introduce phase mismatches. The phase error of the $n$-th element is modeled as $\delta_{\varphi_n} \sim \mathcal{N_{\mathrm{t}}}(0,\sigma_\mathsf{p}^2)$. The impaired array response becomes
    \begin{equation}
        \vspace{-5pt}
        \mathbf{a}_{\mathrm{phase}}(N_{\mathrm{t}},\vartheta) = \boldsymbol{\Phi}\,\mathbf{a}(N_{\mathrm{t}},\vartheta),
    \end{equation}
    where
    \begin{equation}
        \vspace{-5pt}
        \boldsymbol{\Phi} = \operatorname{diag}\!\left(e^{j\delta_{\varphi_1}},\; e^{j\delta_{\varphi_2}},\; \ldots,\; e^{j\delta_{\varphi_N}}\right).
    \end{equation}
    The diagonal matrix $\boldsymbol{\Phi}$ characterizes the random phase offsets across array elements.
\end{itemize}

\begin{figure}[tbp!]
    \centering
    \includegraphics[width=0.4\textwidth]{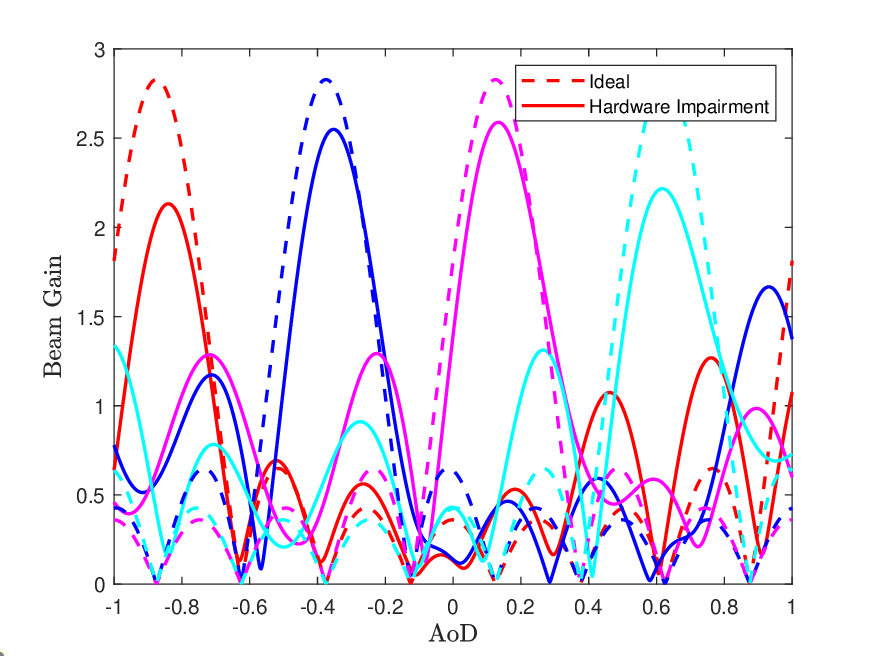}
    \vspace{-5pt}
    \caption{Comparison between the ideal and distorted beam patterns from the DFT codebook with $N_{\mathrm{t}} = 8$, specifically $\{\boldsymbol{f}_{8,1}, \boldsymbol{f}_{8,3}, \boldsymbol{f}_{8,5}, \boldsymbol{f}_{8,7}\}$.}
    \label{fig_DFT_Codebook_Impair}
    \vspace{-10pt}
\end{figure}

The position error is dependent on the incident angle, whereas the phase error is independent of it. Both impairments cause the actual array response $\mathbf{a}_{\mathrm{imp}}(N_{\mathrm{t}},\vartheta)$ to deviate from the ideal response $\mathbf{a}(N_{\mathrm{t}},\vartheta)$. To better illustrate these distortions, we set $\sigma_\mathsf{d} = 0.05\lambda$ and $\sigma_\mathsf{p} = 0.1\pi$, and plot $\{\boldsymbol{f}_{8,1}, \boldsymbol{f}_{8,3}, \boldsymbol{f}_{8,5}, \boldsymbol{f}_{8,7}\}$ as Fig.~\ref{fig_DFT_Codebook_Impair}. As shown in this figure, these impairments introduce significant distortions to the ideal beam patterns, and consequently to the beam gain ratio. This degradation directly affects the performance of both the proposed QSSR method and conventional codebook-based approaches, highlighting the importance of antenna calibration.

\subsection{Proposed Self-Calibration Methods}

To mitigate the degradation caused by hardware impairments, we propose an online self-calibration framework that integrates parametric compensation into the QSSR-Net beam alignment process. The main idea is to represent the effects of impairments through a set of trainable calibration parameters, which are iteratively refined during system operation without requiring additional pilot signals or measurement overhead. 

Specifically, to capture these hardware impairments, including position errors and phase errors, we introduce four diagonal learnable matrices: $\hat{\mathbf{D}}_{\mathrm{t}}$ and $\hat{\mathbf{D}}_{\mathrm{r}}$ to model the transmit- and receive-side position errors, and $\hat{\boldsymbol{\Phi}}_{\mathrm{t}}$ and $\hat{\boldsymbol{\Phi}}_{\mathrm{r}}$ to characterize the corresponding phase distortions. To compensate for these impairments, the network is designed to learn physically meaningful element-wise parameters, namely $\{\hat{\delta}_{\mathrm{t}, d_n}, \hat{\delta}_{\mathrm{r}, d_n}, \hat{\delta}_{\mathrm{t}, \varphi_n}, \hat{\delta}_{\mathrm{r}, \varphi_n}\}$. This structured parameterization explicitly embeds prior knowledge of the hardware impairment model, thereby reducing the effective degrees of freedom and accelerating learning convergence. All parameters are initialized to zero, corresponding to the ideal hardware case without impairments.

\begin{algorithm}[tbp!]
    \SetKwInOut{Input}{Input}\SetKwInOut{Output}{Output}
    \let\oldnl\nl
    \newcommand{\nonl}{\renewcommand{\nl}{\let\nl\oldnl}}
    
    \Output{Compensation matrices $\hat{\mathbf{D}}_{\mathrm{r}}$, $\hat{\mathbf{D}}_{\mathrm{t}}$, $\hat{\boldsymbol{\Phi}}_{\mathrm{r}}$, and $\hat{\boldsymbol{\Phi}}_{\mathrm{t}}$}

    \BlankLine
    \textbf{Offline:} Train QSSR-Net on impairment-free simulated channels.\\
    Initialize $\hat{\mathbf{D}}_{\mathrm{r}}$, $\hat{\mathbf{D}}_{\mathrm{t}}$, $\hat{\boldsymbol{\Phi}}_{\mathrm{r}}$, and $\hat{\boldsymbol{\Phi}}_{\mathrm{t}}$ as trainable calibration variables.\\
    
    \BlankLine
    \While{System is operating}{
        Perform beam alignment with current compensation to obtain estimated angles $(\hat{\theta}, \hat{\phi})$ using QSSR-Net, and collect the measured received powers $\mathbf{p}_{\mathrm{t}}^{\star}$ and $\mathbf{p}_{\mathrm{r}}^{\star}$ from the final-layer beams.\\
        
        Reconstruct the virtual channel $\widehat{\mathbf{H}}_{\text{imp}}$ according to (\ref{equ_H_reconstruct}).\\
        
        Synthesize the received powers $\hat{\mathbf{p}}_{\mathrm{t}}^{\star}$ and $\hat{\mathbf{p}}_{\mathrm{r}}^{\star}$ for the final-layer scanned beams based on $\widehat{\mathbf{H}}_{\text{imp}}$.\\
        
        Fix the QSSR-Net weights and update $\{\hat{\mathbf{D}}_{\mathrm{r}}, \hat{\mathbf{D}}_{\mathrm{t}}, \hat{\boldsymbol{\Phi}}_{\mathrm{r}}, \hat{\boldsymbol{\Phi}}_{\mathrm{t}}\}$ by minimizing the calibration loss in (\ref{equ_calibration_loss}).\\
    }
    \caption{QSSR-Net with Parametric Antenna Calibration}
    \label{alg_QSSR_Impair}
\end{algorithm}

The proposed method can be summarized as Algorithm~\ref{alg_QSSR_Impair}. 
First, the QSSR-Net is trained using impairment-free simulated channels, ensuring that the network learns the ideal mapping between beam measurements and AoA/AoD. During deployment, the system performs beam alignment with the current set of compensation parameters, producing estimates $(\hat{\theta}, \hat{\phi})$ via QSSR-Net. The received powers at the final search layer are also recorded as $\mathbf{p}_{\mathrm{t}}^{\star}$ and $\mathbf{p}_{\mathrm{r}}^{\star}$. A virtual channel $\widehat{\mathbf{H}}_{\text{imp}}$ is then reconstructed using the estimated angles and the current compensation parameters as
\begin{equation} 
    \vspace{-2pt}
    \widehat{\mathbf{H}}_{\text{imp}}
    = \hat{\mathbf{D}}_{\mathrm{r}}(\hat{\theta})\,\hat{\boldsymbol{\Phi}}_{\mathrm{r}}\,
    \mathbf{a}(N_{\mathrm{r}},\hat{\theta})
    \mathbf{a}^{H}(N_{\mathrm{t}},\hat{\phi})\,\hat{\boldsymbol{\Phi}}_{\mathrm{t}}^{H}\,
    \hat{\mathbf{D}}_{\mathrm{t}}^{H}(\hat{\phi}).
    \label{equ_H_reconstruct}
    \vspace{-2pt}
\end{equation}
Based on $\widehat{\mathbf{H}}_{\text{imp}}$, the predicted beam measurement powers $\hat{\mathbf{p}}^{\star}_{\mathrm{t}}$ and $\hat{\mathbf{p}}^{\star}_{\mathrm{r}}$ are synthesized. Keep the QSSR-Net weights fixed, and the calibration parameters $\{\hat{\mathbf{D}}_{\mathrm{r}},\hat{\mathbf{D}}_{\mathrm{t}},\hat{\boldsymbol{\Phi}}_{\mathrm{r}},\hat{\boldsymbol{\Phi}}_{\mathrm{t}}\}$ are updated by minimizing the squared error between measured and predicted powers as
\begin{equation} 
        \mathcal{L} 
        =  \left\|\frac{\mathbf{p}_{\mathrm{t}}^{\star}}
        {\max \mathbf{p}_{\mathrm{t}}^{\star}}
        - \frac{\hat{\mathbf{p}}_{\mathrm{t}}^{\star}}
        {\max \hat{\mathbf{p}}_{\mathrm{t}}^{\star}}\right\|^{2}
        + \left\|\frac{\mathbf{p}_{\mathrm{r}}^{\star}}
        {\max \mathbf{p}_{\mathrm{r}}^{\star}}
        - \frac{\hat{\mathbf{p}}_{\mathrm{r}}^{\star}}
        {\max \hat{\mathbf{p}}_{\mathrm{r}}^{\star}}\right\|^{2}. 
    \label{equ_calibration_loss}
\end{equation}

Over prolonged system operation, the antenna array is gradually self-calibrated. The proposed method requires no additional beam measurements or external calibration hardware, relying solely on the outputs of QSSR-Net and the beam measurements already obtained during beam alignment. Furthermore, the calibration parameters can be periodically updated during deployment (rather than after every alignment) to reduce computational overhead, ensuring robustness against time-varying impairments such as temperature drift and hardware aging. By restoring the effective array response closer to the ideal model, the proposed method enhances the reliability of QSSR-Net predictions and improves the accuracy of super-resolution beam alignment. This process creates a positive feedback loop, since more accurate beam alignment further facilitates improved antenna calibration. A theoretical analysis of this process is provided in the \textbf{Appendix}.

\section{Simulation Results}
\label{sec_simulation_result}

In this section, we present numerical results to evaluate the effectiveness of the proposed methods. We first describe the simulation parameters, and then sequentially compare the proposed algorithms with several benchmarks to demonstrate the advantages.

\subsection{Simulation Details}

Unless otherwise specified, we assume that the transmitter is equipped with $N_{\mathrm{t}} = 64$ antennas and the receiver has a $N_{\mathrm{r}} = 16$ antennas. The channel follows the S-V model in (\ref{equ_SV_model}) with $L = 3$ resolvable paths. The AoDs and AoAs are independently drawn from a uniform distribution over $[-1,1]$. The complex path gains are generated as $\alpha_1 \sim \mathcal{CN}(0,1)$ for the dominant LoS path, while the non-line-of-sight (NLoS) paths follow $\alpha_2, \alpha_3 \sim \mathcal{CN}(0,0.01)$. This setting reflects a typical LoS-dominated propagation environment. The channels are normalized to $\|\mathbf{H}\|^2 = 1$, and the SNR is defined as $\operatorname{SNR} = \|\mathbf{H}\|^2 /\sigma^2$.

To further validate the robustness of the proposed methods, we evaluate their performance on both a complex NLoS scenario dataset and the public DeepMIMO dataset \cite{alkhateebDeepMIMOGenericDeep2019}. For the NLoS scenario, the channel follows the S-V model in (\ref{equ_SV_model}) with $L = 8$ resolvable paths, and the complex gains of all paths are modeled as $\mathcal{CN}(0,1)$. For the DeepMIMO dataset, we consider the \emph{Boston scenario}. This dataset is generated using accurate ray-tracing simulations from Wireless InSite, which capture realistic propagation characteristics in an urban environment. The system operates at a carrier frequency of $f_c = 28$ GHz. User locations are distributed uniformly within the shaded area among the buildings, encompassing both LoS and NLoS conditions. For each channel realization, up to eight multipath components are simulated. 

The proposed QSSR-based method employs a GRU network trained on simulated LoS-dominated channel realizations over an SNR range of $[5,30]$ dB, assuming ideal hardware during training. The hidden dimensions of the GRU as well as the fully connected layers are set to 64. The conventional Adam optimizer is adopted with a step-based learning rate schedule, where the learning rate is multiplied by 0.95 every 10 training epochs. The batch size is set to 100.

For performance evaluation, we compare our method against three representative benchmarks:
\begin{itemize}
    \item \textbf{Binary Search:} Both the transmitter and receiver perform binary beam alignment over the DFT codebook, following a procedure similar to the proposed QSSR algorithm but without the super-resolution process.    
    \item \textbf{Exhaustive Search:} The transmitter and receiver exhaustively evaluate all possible beam pairs in finest codebooks. This approach serves as an upper bound for codebook-based alignment without super-resolution.
    \item \textbf{SR-BSNet}~\cite{fanSuperResolutionBasedBeam2022}: An image super-resolution baseline that reconstructs a fine-grained $N_{\mathrm{t}} \times N_{\mathrm{r}}$ power map from coarse wide-beam measurements of size $N_{\mathrm{t}}/4 \times N_{\mathrm{r}}/4$.
\end{itemize}

\subsection{Beam Alignment Overhead}

\begin{figure}[tb!]
    \centering
    \includegraphics[width=0.4\textwidth, clip]{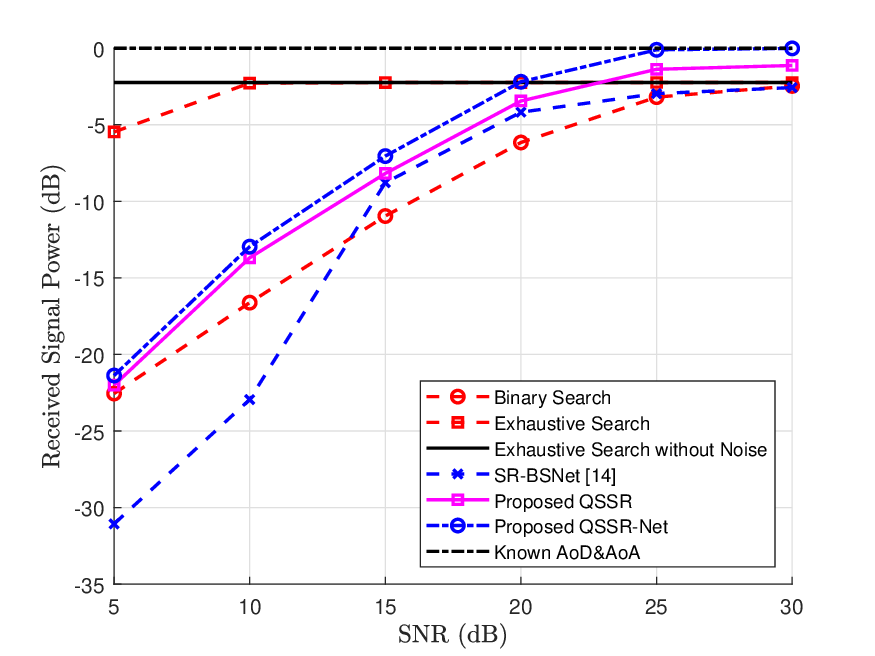}
    \vspace{-5pt}
    \caption{Average received power vs. SNR for various beam alignment schemes.}
    \label{fig_SV_SNR}
    \vspace{-8pt}
\end{figure}

\begin{figure}[tb!]
    \centering
    \includegraphics[width=0.4\textwidth, clip]{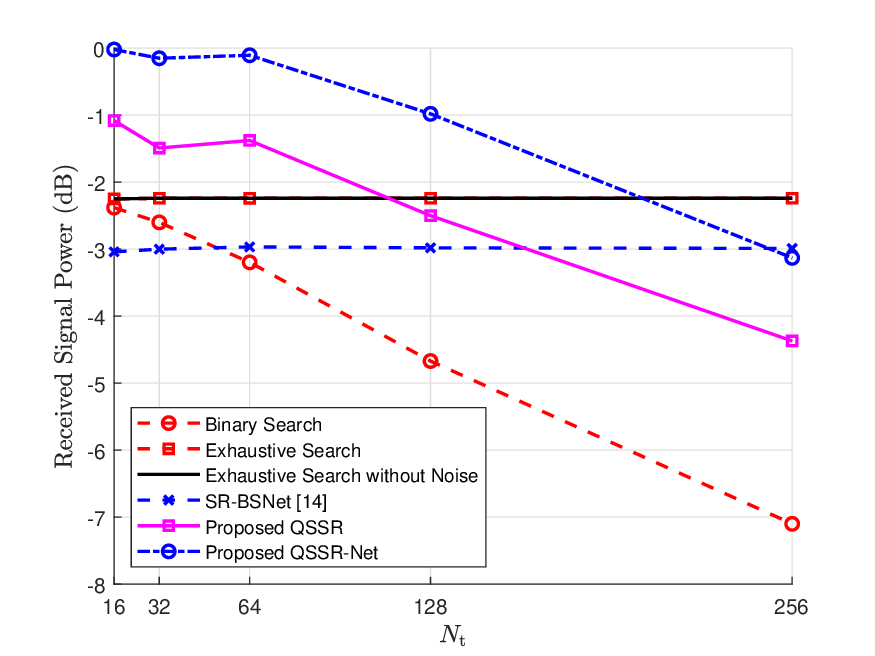}
    \vspace{-5pt}
    \caption{Average received power vs. number of transmit antennas for various beam alignment schemes.}
    \label{fig_SV_Antenna}
    \vspace{-8pt}
\end{figure}

We first analyze the required beam measurement overhead of the considered methods. The initial exhaustive search method scans all possible narrow beam pairs in the codebooks, resulting in a total of $N_{\mathrm{r}} \times N_{\mathrm{t}} = 16 \times 64 = 1024$ beam measurements. The binary search method requires $2\left\lceil \log_2 N_{\mathrm{r}} \right\rceil + 2\left\lceil \log_2 N_{\mathrm{t}} \right\rceil = 20$ beam measurements, which significantly reduce the overhead compared to exhaustive search. This highlights the advantage of hierarchical search. The SR-BSNet method scans $N_{\mathrm{r}}/4 \times N_{\mathrm{t}}/4 = 16/4 \times 64/4 = 64$ wide beams to obtain coarse measurements. The proposed QSSR method probes $ 4\left\lceil \log_4 N_{\mathrm{r}} \right\rceil + 4\left\lceil \log_4 N_{\mathrm{t}} \right\rceil = 20$ beam pairs in total, which matches the overhead of binary search but enables super-resolution estimation of continuous AoA and AoD. The proposed QSSR-Net method shares the same measurement overhead as QSSR, since it is built upon the QSSR framework. 

\vspace{-10pt}
\subsection{Performance versus SNR}

We next analyze the received power performance of the considered methods across different SNR values, as shown in Fig.~\ref{fig_SV_SNR}. The results show that the received signal power of all methods increases with SNR. Among them, the DL-based SR-BSNet achieves performance comparable to binary search in the high-SNR regime but requires significantly higher measurement overhead and suffers noticeable degradation at low-to-moderate SNRs due to its sensitivity to noise in the reconstruction process. In contrast, the proposed QSSR algorithm consistently outperforms binary search across the entire SNR range, validating the effectiveness of the super-resolution principle. Furthermore, by employing a GRU to learn the inverse mapping from beam measurements to angular information, QSSR-Net further enhances performance. Remarkably, in the high-SNR regime (SNR $>20$ dB), QSSR-Net even surpasses exhaustive search, which is commonly regarded as the upper bound of codebook-based alignment. This is because the performance of exhaustive search is inherently limited by the angular quantization of the DFT codebook. Moreover, QSSR-Net approaches the performance of the ``Known AoD\&AoA'' benchmark, which ideally steers the beams toward the true angle of the dominant path, when the SNR exceeds 20~dB. These results demonstrate that the proposed algorithms not only retain the same beam training overhead as binary search but also achieve a superior trade-off between efficiency and accuracy by breaking through the resolution barrier of conventional codebook-based alignment.

\subsection{Performance versus Number of Antennas}

We next analyze the received power performance of the considered methods with different numbers of transmit antennas, as shown in Fig.~\ref{fig_SV_Antenna}. For a fair comparison, the number of receive antennas is fixed at $N_{\mathrm{r}} = 16$, while the number of transmit antennas is varied. The SNR is set to 25~dB. From the figure, it can be observed that the performance of exhaustive search and SR-BSNet remains essentially constant as $N_{\mathrm{t}}$ increases. This is because these methods proportionally increase the number of beam scans to maintain their performance. In contrast, the performance of binary search and the proposed QSSR method degrades with larger $N_{\mathrm{t}}$, since a larger array size also increases the number of codebook layers that must be scanned. Each additional layer introduces the possibility of incorrect codeword selection, and the probability of error accumulates across layers. An interesting observation is that the case of $N_{\mathrm{t}} = 32$ yields slightly worse performance than $N_{\mathrm{t}} = 64$, which deviates from the general trend. This behavior arises because with $N_{\mathrm{t}} = 32$, the algorithm scans only $\mathcal{W}_{4}$ and $\mathcal{W}_{16}$, whereas with $N_{\mathrm{t}} = 64$, an additional scan over $\mathcal{W}_{64}$ is performed. The extra resolution layer enables QSSR-Net to better exploit the super-resolution principle, thereby improving angle estimation accuracy and enhancing overall performance.

\begin{figure}[tb!]
    \centering
    \includegraphics[width=0.4\textwidth, clip]{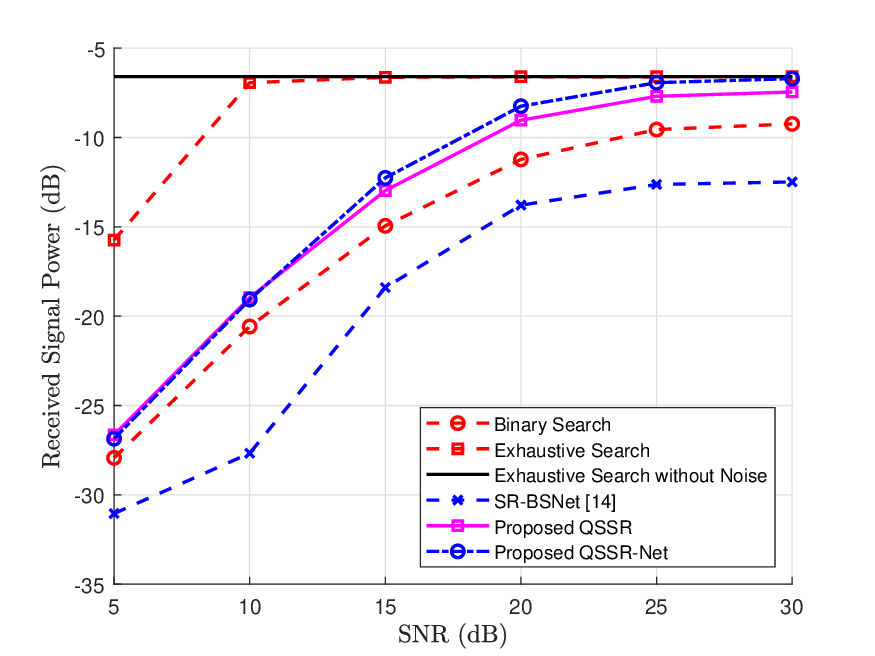}
    \vspace{-5pt}
    \caption{Average received power vs. SNR for various beam alignment schemes in NLoS scenarios.}
    \vspace{-8pt}
    \label{fig_SV_NLoS_SNR}
\end{figure}

\begin{figure}[tb!]
    \centering
    \includegraphics[width=0.4\textwidth, clip]{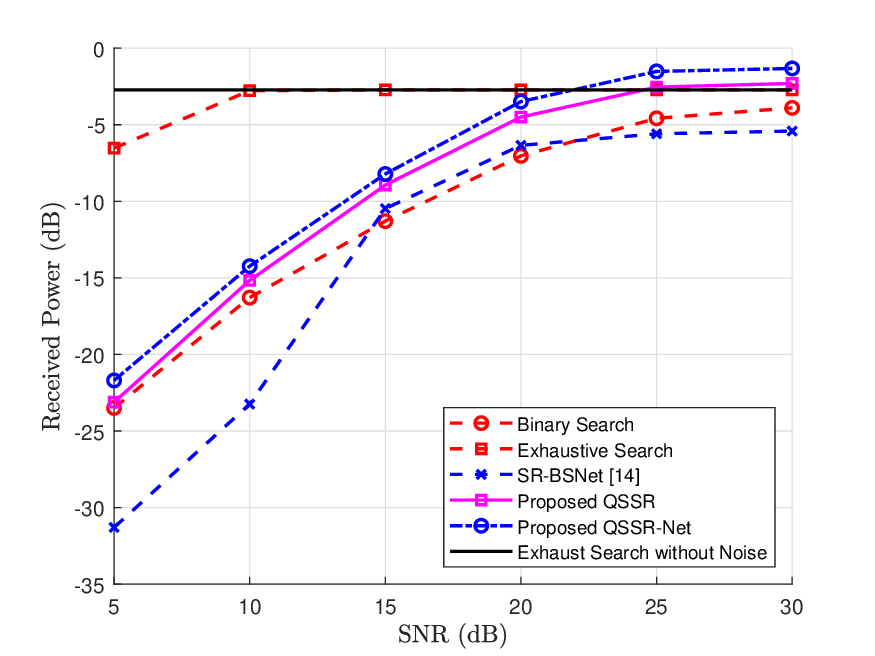}
    \vspace{-5pt}
    \caption{Average received power vs. SNR for various beam alignment schemes in DeepMIMO dataset.}
    \vspace{-8pt}
    \label{fig_DeepMIMO_O1_SNR}
\end{figure}

\subsection{Robustness Analysis}

\begin{figure*}[tb!]
    \centering
    \subfigure[Binary Search]{
    \includegraphics[width=0.36\textwidth]{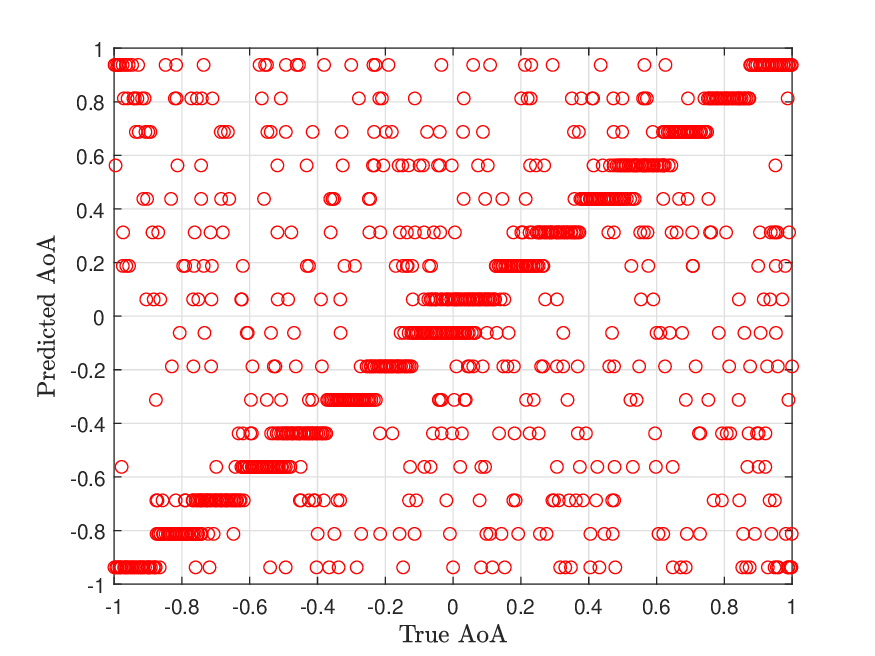} 
    }    
    \vspace{-10pt}
    \subfigure[SR-BSNet]{
    \includegraphics[width=0.36\textwidth]{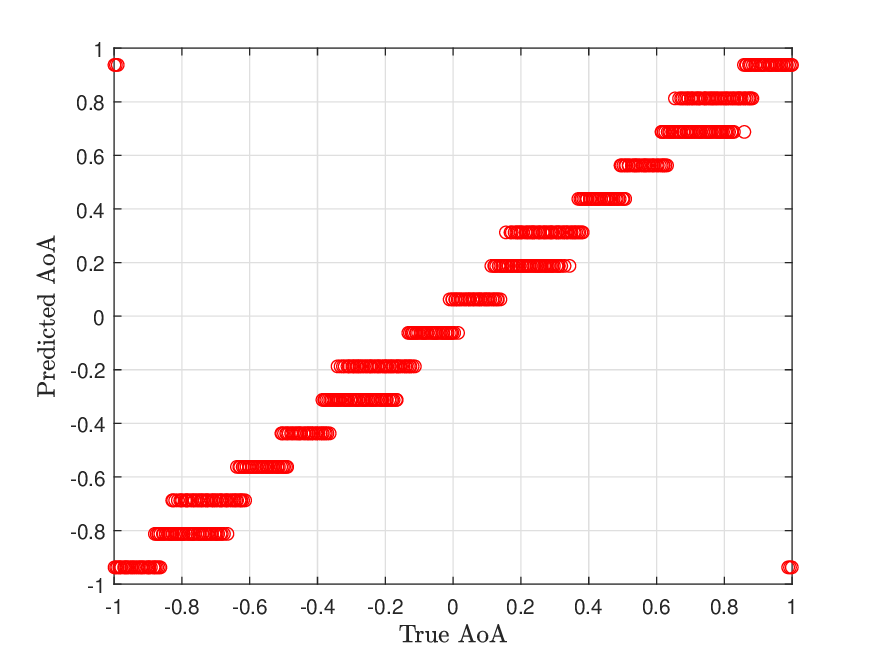} 
    }
    \subfigure[Proposed QSSR]{
    \includegraphics[width=0.36\textwidth]{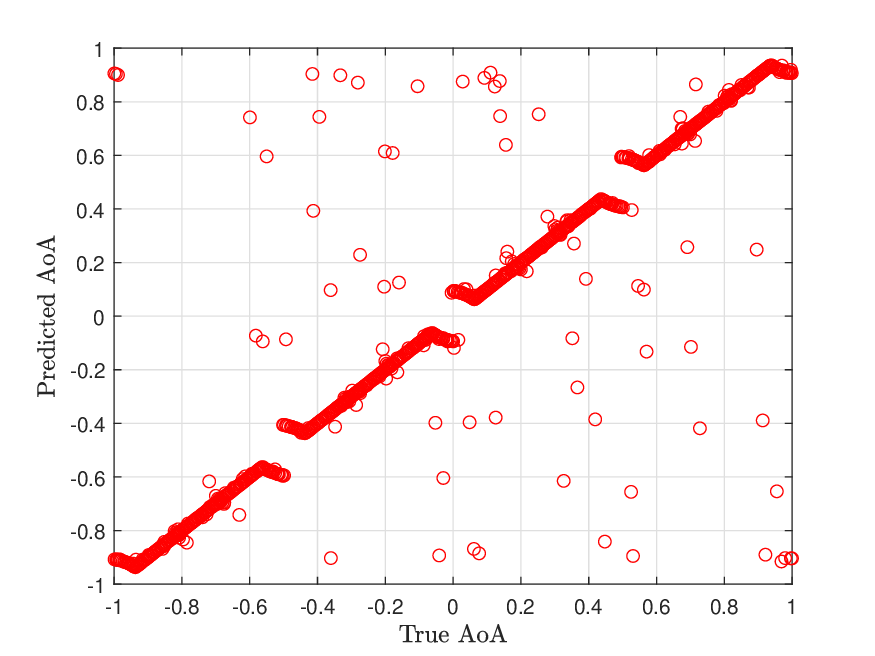} 
    }
    \subfigure[Proposed QSSR-Net]{
    \includegraphics[width=0.36\textwidth]{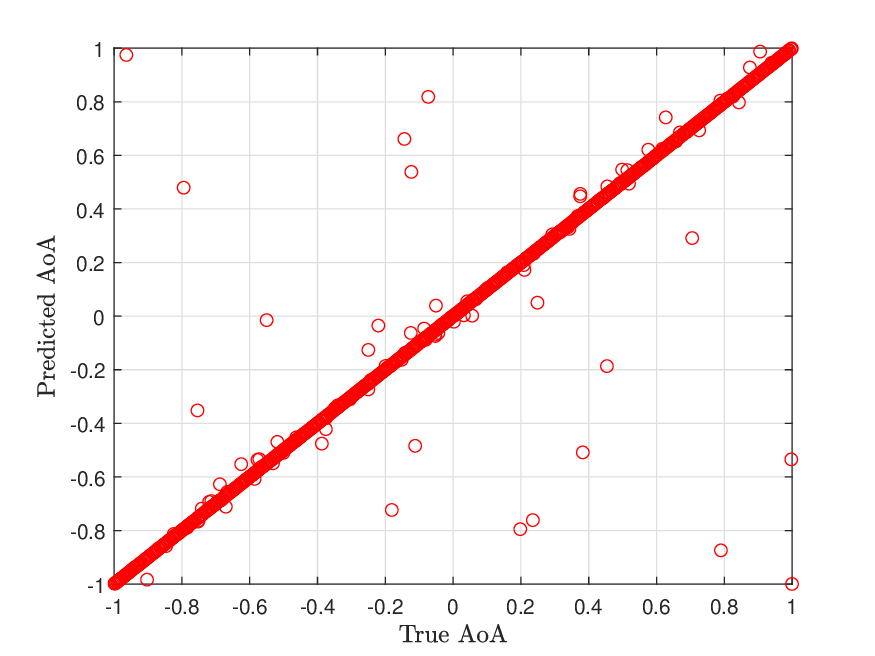} 
    }
    \caption{True vs. predicted AoA of different algorithms.}
    \vspace{-10pt}
    \label{fig_error_analysis}
\end{figure*}

To validate the robustness of the proposed QSSR-Net, we first train the network using LoS channels generated by the S-V model and then directly deploy it on a complex NLoS dataset. This setup emulates environmental changes that alter channel statistics and typically degrade the performance of conventional DL-based methods. The received signal power across different SNR values is shown in Fig.~\ref{fig_SV_NLoS_SNR}. Because the power is distributed among multiple paths, the effective power of the dominant path is reduced, leading to performance degradation. As observed from the figure, SR-BSNet suffers significant degradation in this complex propagation environment. Since SR-BSNet relies on coarse-to-fine power reconstruction, a network trained on LoS channels may focus only on a single path even in NLoS scenarios, resulting in poor generalization and performance loss, in some cases performing worse than binary search. In contrast, QSSR-Net consistently outperforms binary search. This robustness arises because the proposed QSSR algorithm and QSSR-Net are built upon hierarchical search, where the final-layer beams typically cover only one path, thereby mitigating multipath interference. Furthermore, QSSR-Net learns the inverse mapping from normalized received power measurements to angular information, enhancing its generalization capability. 

We further validate robustness by deploying QSSR-Net on the DeepMIMO dataset. The received signal power across different SNR values is shown in Fig.~\ref{fig_DeepMIMO_O1_SNR}. The DeepMIMO dataset represents a realistic propagation environment with multiple paths, blockages, and diffuse scattering. Its channels are generated from accurate ray-tracing simulations, which differ substantially from the S-V model used for training. Even under these conditions, QSSR-Net demonstrates strong robustness, outperforming binary search and surpassing exhaustive search at high SNRs. These results highlight the practical applicability of the proposed method in real-world scenarios.

\subsection{Error Analysis}

To further investigate angular estimation accuracy, Fig.~\ref{fig_error_analysis} compares the true versus predicted AoAs under different beam alignment schemes when SNR $= 25$ dB. The results highlight several distinctive error patterns. Binary search provides only grid-based estimates restricted to the codebook, resulting in coarse resolution and pronounced quantization errors. SR-BSNet exhibits wider horizontal error bands, indicating larger estimation variance and frequent confusion between neighboring codewords due to its reliance on coarse-to-fine power reconstruction. In contrast, the proposed QSSR significantly improves the resolution by exploiting the monotonic power ratio property, but relatively large errors still occur near the boundaries of beam sectors. This is because when the true AoA lies at a sector boundary, the auxiliary beam is likely to be incorrectly selected. By incorporating DL, QSSR-Net effectively resolves these boundary ambiguities by learning to correct nonlinear distortions in the ratio–angle mapping, thereby achieving highly accurate AoA estimates across the entire angular domain. These observations demonstrate the clear advantage of the proposed DL-empowered framework in providing fine-grained and robust beam alignment.

\begin{figure} 
    \centering
    \includegraphics[width=0.4\textwidth]{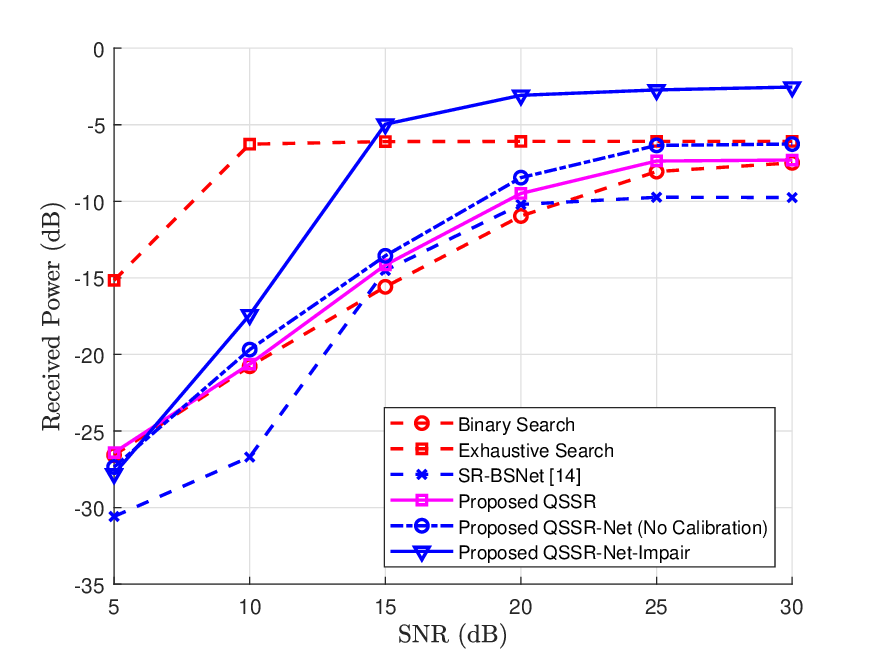}
    \vspace{-5pt}
    \caption{Average received power vs. SNR for various beam alignment schemes under hardware impairments.}
    \vspace{-5pt}
    \label{fig_Impairments}
\end{figure}

\begin{figure} 
    \centering
    \includegraphics[width=0.4\textwidth]{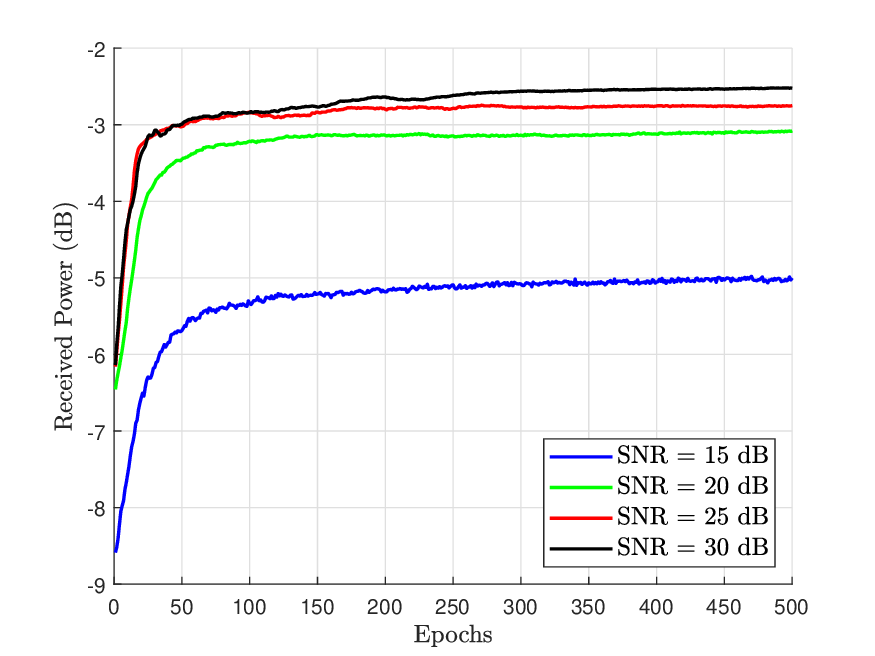}
    \vspace{-5pt}
    \caption{Average received power vs. self-calibration training epochs for different SNR values.}
    \label{fig_Impairments_epoch}
    \vspace{-5pt}
\end{figure}

\subsection{Performance under Hardware Impairments}

To simulate hardware impairments, we primarily consider position errors and phase errors, while noting that other types of impairments can be compensated in a similar manner. Specifically, we set $\sigma_\mathsf{d} = 0.05\lambda$ and $\sigma_\mathsf{p} = 0.1\pi$. The calibration is performed at a specific SNR. For example, QSSR-Net-Impair tested at $\mathrm{SNR}=30$ dB is also calibrated at $\mathrm{SNR}=30$ dB. This setting reflects practical deployment, where self-calibration is conducted under the same SNR conditions as those experienced during operation. 

The performance of different algorithms versus SNR is shown in Fig.~\ref{fig_Impairments}. As observed, all methods degrade under hardware impairments, since distorted beam patterns cause erroneous beam selection and inaccurate angle estimation. The performance of the QSSR algorithm drops to a level comparable to binary search, because beam distortions corrupt the gain ratios that directly determine angle estimation, leading to significant performance loss. Nevertheless, QSSR-Net still approaches the performance of exhaustive search in the high-SNR regime, demonstrating strong robustness. Furthermore, for $\mathrm{SNR}>5$ dB, incorporating the proposed self-calibration method enables QSSR-Net-Impair to effectively mitigate impairment-induced degradation. At $\mathrm{SNR}=5$ dB, its performance slightly deteriorates, as noisy angle estimates yield inaccurate channel reconstruction, which in turn hampers calibration. When $\mathrm{SNR}>15$ dB, QSSR-Net-Impair even surpasses exhaustive search, highlighting the effectiveness of the proposed online self-calibration framework in practical scenarios.

We also compare the performance during the self-calibration process. The received power over different SNRs during calibration is plotted in Fig.~\ref{fig_Impairments_epoch}. The performance of QSSR-Net-Impair improves progressively with the number of training epochs, as the antenna array is gradually calibrated. Notably, QSSR-Net-Impair at higher SNRs converges to better performance due to reduced measurement noise. Overall, the calibration process converges within 300 epochs, demonstrating the efficiency of the proposed method.

\section{Conclusion}
\label{sec_conclusion}

In this study, we investigated beam alignment in mmWave massive MIMO systems. We first analyzed the hierarchical codebook structure and identified the monotonic property of the beam gain ratio between two codewords in the same layer. These insights guided the algorithm design. Specifically, we proposed the QSSR algorithm, which combines the advantages of super-resolution estimation and hierarchical search. We further extended this approach by integrating a GRU network to learn the inverse mapping from multiple beam measurements to angular information, thereby enhancing estimation accuracy. To address the impact of hardware impairments, we proposed the QSSR-Net-Impair algorithm, which incorporates an online self-calibration mechanism. Simulation results demonstrate that the proposed methods significantly outperform existing schemes, achieving up to 2.2~dB performance gain due to the super-resolution property, while maintaining logarithmic beam training overhead suitable for large arrays. Moreover, QSSR-Net exhibits strong robustness in both simulated channels and ray-tracing-based datasets. Finally, the proposed self-calibration method effectively compensates for distortions in the array response, enhancing the reliability of angle estimation and improving overall beam alignment performance by approximately 4~dB.

\section*{Appendix: Theoretical Analysis of Self-Calibration}

The effectiveness of the proposed self-calibration framework can be explained by formulating it as a parameter estimation problem. To simplify the analysis, we assume a pure LoS channel with $L=1$, while ignoring the effects of NLoS components and noise. Under this setting, the effective channel can be expressed as
\begin{equation} 
    \mathbf{H}_{\text{imp}} = \alpha \mathbf{D}_{\mathrm{r}}(\theta)\,\boldsymbol{\Phi}_{\mathrm{r}}\,\mathbf{a}(N_{\mathrm{r}},\theta)\,
    \,\mathbf{a}^H(N_{\mathrm{t}},\phi)\,\boldsymbol{\Phi}_{\mathrm{t}}^H\,\mathbf{D}_{\mathrm{t}}^H(\phi),
\end{equation}
where $\alpha$ denotes the complex path gain, and $\theta$ and $\phi$ represent the AoA and AoD, respectively. Consider the transmitter side during the final-layer probing in the beam alignment process. The conjugate transposes of $\{\hat{\mathbf{D}}_{\mathrm{r}}, \hat{\mathbf{D}}_{\mathrm{t}}, \hat{\boldsymbol{\Phi}}_{\mathrm{r}}, \hat{\boldsymbol{\Phi}}_{\mathrm{t}}\}$ are applied to compensate for antenna impairments. For convenience, let $\hat{\mathbf{f}}_i = \hat{\boldsymbol{\Phi}}_{\mathrm{t}} \, \hat{\mathbf{D}}_{\mathrm{t}}(\bar{\phi}_i) \bar{\mathbf{f}}_i$ and $\hat{\mathbf{w}} = \hat{\boldsymbol{\Phi}}_{\mathrm{r}} \, \hat{\mathbf{D}}_{\mathrm{r}}(\hat{\theta}) \bar{\mathbf{w}}$. Here, $\bar{\mathbf{f}}_i$ and $\bar{\phi}_i$ are the $i$-th probed beamforming vector and its corresponding steering angle, respectively, while $\bar{\mathbf{w}}$ is the fixed combining vector at the receiver. The steering angle of $\hat{\mathbf{w}}$ is denoted as $\hat{\theta}$. 

The received signal powers corresponding to the four scanned beams can be represented as
\begin{equation}     
        \mathbf{p}_{\mathrm{t}}^{\star} = \Big[|\hat{\mathbf{w}}^H  \mathbf{H}_\text{imp} \hat{\mathbf{f}}_1|^2, \ldots, |\hat{\mathbf{w}}^H \mathbf{H}_\text{imp} \hat{\mathbf{f}}_4|^2 \Big]^T.
\end{equation}
Without loss of generality, assume that $\mathbf{p}_{\mathrm{t}}[1]$ corresponds to the maximum received power. The normalized measured power vector is then given by  
\begin{align}      
        \mathbf{p}_\text{nor} & = \frac{\mathbf{p}_{\mathrm{t}}^{\star}}{\max \mathbf{p}_{\mathrm{t}}^{\star}} = \Big[1, \frac{\mathbf{p}_{\mathrm{t}}^{\star} [2]}{\mathbf{p}_{\mathrm{t}}^{\star} [1]}, \frac{\mathbf{p}_{\mathrm{t}}^{\star} [3]}{\mathbf{p}_{\mathrm{t}}^{\star} [1]}, \frac{\mathbf{p}_{\mathrm{t}}^{\star} [4]}{\mathbf{p}_{\mathrm{t}}^{\star} [1]} \Big]^T \notag\\
        & = \Bigg[1, \frac{|\mathbf{a}^H(N_{\mathrm{t}},\phi) \boldsymbol{\Phi}_{\mathrm{t}}^H {\mathbf{D}}_{\mathrm{t}}^H(\phi) \hat{\mathbf{f}}_2|^2}{|\mathbf{a}^H(N_{\mathrm{t}},\phi) \boldsymbol{\Phi}_{\mathrm{t}}^H  \mathbf{D}_{\mathrm{t}}^H(\phi) \hat{\mathbf{f}}_1|^2}, \ldots, \notag  \\
        & \quad\quad\quad\, \frac{|\mathbf{a}^H(N_{\mathrm{t}},\phi) \boldsymbol{\Phi}_{\mathrm{t}}^H {\mathbf{D}}_{\mathrm{t}}^H(\phi) \hat{\mathbf{f}}_4|^2}{|\mathbf{a}^H(N_{\mathrm{t}},\phi) \boldsymbol{\Phi}_{\mathrm{t}}^H  \mathbf{D}_{\mathrm{t}}^H(\phi) \hat{\mathbf{f}}_1|^2} \Bigg]^T,
    \label{equ_p_nor}
\end{align}

After beam alignment, the estimated AoA/AoD $(\hat{\theta}, \hat{\phi})$ are obtained via QSSR-Net. Based on these estimates and the current compensation parameters $\{\mathbf{D}_{\mathrm{r}}, \mathbf{D}_{\mathrm{t}}, \boldsymbol{\Phi}_{\mathrm{r}}, \boldsymbol{\Phi}_{\mathrm{t}}\}$, a virtual channel $\widehat{\mathbf{H}}_\text{imp}$ can be reconstructed as in (\ref{equ_H_reconstruct}). The corresponding normalized synthesized received power vector is expressed as  
\begin{align}        
        \hat{\mathbf{p}}_\text{nor}  & = \Bigg[1, \frac{|\mathbf{a}^H(N_{\mathrm{t}},\hat{\phi}) \hat{\boldsymbol{\Phi}}_{\mathrm{t}}^H \hat{\mathbf{D}}_{\mathrm{t}}^H(\hat{\phi}) \hat{\mathbf{f}}_2|^2}{|\mathbf{a}^H(N_{\mathrm{t}},\hat{\phi}) \hat{\boldsymbol{\Phi}}_{\mathrm{t}}^H  \hat{\mathbf{D}}_{\mathrm{t}}^H(\hat{\phi}) \hat{\mathbf{f}}_1|^2}, \ldots, \notag \\
        & \quad\quad\quad\, \frac{|\mathbf{a}^H(N_{\mathrm{t}},\hat{\phi}) \hat{\boldsymbol{\Phi}}_{\mathrm{t}}^H \hat{\mathbf{D}}_{\mathrm{t}}^H(\hat{\phi}) \hat{\mathbf{f}}_4|^2}{|\mathbf{a}^H(N_{\mathrm{t}},\hat{\phi}) \hat{\boldsymbol{\Phi}}_{\mathrm{t}}^H  \hat{\mathbf{D}}_{\mathrm{t}}^H(\hat{\phi}) \hat{\mathbf{f}}_1|^2} \Bigg]^T.
   \label{equ_p_hat_nor}
\end{align}
The calibration procedure for the transmitter is thus equivalent to minimizing the discrepancy between the measured and synthesized power vectors.

Assuming that when the compensation parameters perfectly match the true impairments, the estimation of QSSR-Net is accurate, which has been validated in the simulations. Thus, when the loss function reaches the minimum value of zero, we can derive that $\mathbf{p}_{\text{nor}} = \hat{\mathbf{p}}_{\text{nor}}$. According to the assumption, the angle estimation is accurate, i.e., $\hat{\phi} = \phi$. Considering the $k$-th element of $\mathbf{p}_{\text{nor}}$, we obtain
\begin{equation}
        \frac{\left|\mathbf{a}^{H}\left(N_{\mathrm{t}}, \phi\right) 
        \boldsymbol{\Phi}_{\mathrm{t}}^{H} \mathbf{D}_{\mathrm{t}}^{H}(\phi) 
        \hat{\mathbf{f}}_{k}\right|^{2}}
        {\left|\mathbf{a}^{H}\left(N_{\mathrm{t}}, \phi\right) 
        \boldsymbol{\Phi}_{\mathrm{t}}^{H} \mathbf{D}_{\mathrm{t}}^{H}(\phi) 
        \hat{\mathbf{f}}_{1}\right|^{2}} 
        = 
        \frac{\left|\mathbf{a}^{H}\left(N_{\mathrm{t}}, \phi\right) 
        \hat{\boldsymbol{\Phi}}_{\mathrm{t}}^{H} \hat{\mathbf{D}}_{\mathrm{t}}^{H}(\hat{\phi}) 
        \hat{\mathbf{f}}_{k}\right|^{2}}
        {\left|\mathbf{a}^{H}\left(N_{\mathrm{t}}, \phi\right) 
        \hat{\boldsymbol{\Phi}}_{\mathrm{t}}^{H} \hat{\mathbf{D}}_{\mathrm{t}}^{H}(\hat{\phi}) 
        \hat{\mathbf{f}}_{1}\right|^{2}} .
\end{equation}

To make this equality hold for every $k$, one sufficient condition is that the measured and synthesized normalized received powers should be proportional. 
Define $c(\phi)$ as the proportionality ratio, then we have
\begin{equation}
    \mathbf{a}^{H}(N_{\mathrm{t}}, \phi) 
    \boldsymbol{\Phi}_{\mathrm{t}}^{H} \mathbf{D}_{\mathrm{t}}^{H}(\phi) 
    = c(\phi)\, 
    \mathbf{a}^{H}(N_{\mathrm{t}}, \phi) 
    \hat{\boldsymbol{\Phi}}_{\mathrm{t}}^{H} \hat{\mathbf{D}}_{\mathrm{t}}^{H}(\hat{\phi}).
\end{equation}
Considering that $\boldsymbol{\Phi}_{\mathrm{t}}^{H} \mathbf{D}_{\mathrm{t}}^{H}(\phi)$ is a diagonal matrix, 
$\mathbf{a}^{H}(N_{\mathrm{t}}, \phi)$ can be eliminated from both sides since it is a row vector, yielding
\begin{equation}
    \boldsymbol{\Phi}_{\mathrm{t}}^{H} \mathbf{D}_{\mathrm{t}}^{H}(\phi) 
    = c(\phi)\, \hat{\boldsymbol{\Phi}}_{\mathrm{t}}^{H} 
    \hat{\mathbf{D}}_{\mathrm{t}}^{H}(\hat{\phi}).
\end{equation}
Taking the $n$-th element on both sides, we obtain
\begin{equation}
    c^{*}(\phi)
    = e^{j\left(\hat{\delta}_{\varphi_{n}}-\delta_{\varphi_{n}}\right)} 
    \cdot e^{j \frac{2 \pi}{\lambda}\left(\hat{\delta}_{d_{n}}-\delta_{d_{n}}\right) \phi}.
\end{equation}

Assuming enough beam measurements are collected at different angles $\phi$, we obtain a series of equations. Since the left-hand side is only related to $\phi$ while the right-hand side is related to $n$, the only possibility is that both sides equal a constant, i.e., $c^{*}(\phi) = c^{*}$. This implies $\hat{\delta}_{\varphi_{n}} - \delta_{\varphi_{n}}$ and $\hat{\delta}_{d_{n}} - \delta_{d_{n}}$ are constants independent of $n$. This indicates that the compensation parameters can converge to the true impairments plus a constant offset. However, the influence of a constant offset can be regarded as a slight translation or rotation of the antenna array, which has negligible impact on the beam pattern. The compensation for the receiver can be similarly derived. 

It should also be noted that there exists a possibility where an incorrect compensation matrix together with a wrong angle estimation may also make the objective function reach zero. However, the self-calibration is a dynamic process. The principle can be interpreted as follows. First, the mapping from impairment parameters to the measured power vector is injective in the considered domain, since distinct impairment realizations lead to distinguishable distortions in $\mathbf{p}_\text{nor}$. Second, as the calibration parameters converge, the synthesized channel $\widehat{\mathbf{H}}_{\text{imp}}$ approaches the true impaired channel $\mathbf{H}_\text{imp}$, thereby restoring the effective array response. Finally, improved calibration enhances the accuracy of QSSR-Net angle predictions, which in turn refines the channel reconstruction and further accelerates calibration. This creates a positive feedback loop, ensuring convergence to an accurate compensation of the hardware impairments. In practice, this process may not always work perfectly due to the presence of noise and multi-path components. Nevertheless, its effectiveness has been validated in the simulation results, where performance improvement is achieved for SNR $\geq$ 10 dB.

\end{document}